\documentclass[twocolumn]{aastex62}
\usepackage{graphicx}
\usepackage{natbib}
\usepackage{amsmath}
\usepackage{tgcursor}
\usepackage{booktabs}
\usepackage{placeins}

\setcitestyle{notesep={ }}

\begin{document}

\title{Oxygen atom reactions with C$_2$H$_6$, C$_2$H$_4$, and C$_2$H$_2$ in ices}

\author{Jennifer B. Bergner}
\affiliation{Harvard University Department of Chemistry and Chemical Biology, Cambridge, MA 02138, USA}

\author{Karin I. \"Oberg}
\affiliation{Harvard-Smithsonian Center for Astrophysics, Cambridge, MA 02138, USA}

\author{Mahesh Rajappan}
\affiliation{Harvard-Smithsonian Center for Astrophysics, Cambridge, MA 02138, USA}

\begin{abstract}
Oxygen atom addition and insertion reactions may provide a pathway to chemical complexity in ices that are too cold for radicals to diffuse and react.  We have studied the ice-phase reactions of photo-produced oxygen atoms with C$_2$ hydrocarbons under ISM-like conditions.  The main products of oxygen atom reactions with ethane are ethanol and acetaldehyde; with ethylene are ethylene oxide and acetaldehyde; and with acetylene is ketene.  The derived branching ratio from ethane to ethanol is $\sim$0.74 and from ethylene to ethylene oxide is $\sim$0.47.  For all three hydrocarbons there is evidence of an effectively barrierless reaction with O($^1$D) to form oxygen-bearing organic products; in the case of ethylene, there may be an additional barriered contribution of the ground-state O($^3$P) atom.  Thus, oxygen atom reactions with saturated and unsaturated hydrocarbons are a promising pathway to chemical complexity even at very low temperatures where the diffusion of radical species is thermally inaccessible.
\end{abstract}


\section{Introduction}
\label{sec:intro}
The inventory of organic molecules present during planet formation is a topic of great interest to origins of life studies.  This inventory is set by a combination of in situ formation in protoplanetary disks and inheritance from the earlier stages of star formation.  Complex organic molecules (COMs), defined as hydrogen-rich organics of 6 or more atoms, have been detected at all stages of star formation, from pre-stellar cores \citep{Oberg2010, Bacmann2012} to protostars \citep{Blake1987, vanDishoeck1995} to protoplanetary disks \citep{Oberg2015, Walsh2016, Favre2018}.  Astrochemical models can reproduce the formation of COMs in lukewarm and hot astrophysical environments ($>$30 K) by using a radical recombination mechanism active in icy grain mantles, in which energetic processing of small molecules is followed by diffusion and recombination to form larger organics, and ice sublimation at warmer ($>$100 K) temperatures \citep[e.g.][]{Garrod2008, Herbst2009}.  Indeed, in laboratory experiments, radical recombination chemistry induced by photolysis \citep[e.g.][]{Gerakines1996, Oberg2009}, radiolysis \citep{Hudson1999, Bennett2005a}, or H atom bombardment \citep{Fedoseev2015, Chuang2016} of simple ice mixtures has been shown to form COMs.  At low temperatures ($\sim$10 K), however, models predict that the diffusion of radical fragments in interstellar ices will be inefficient, and diffusion-limited radical ice chemistry cannot explain the detections of COMs towards very cold sources like pre-stellar cores.  This suggests that the current framework is incomplete, and additional COM formation pathways need to be considered.

A potential low-temperature channel to COM formation is the insertion of an O($^1$D) atom directly into the C-H bond of a hydrocarbon.  Atoms are mobile down to lower temperatures than radical fragments and could therefore still diffuse and react in very cold environments.  The O($^1$D) atom is the first electronically excited state of the oxygen atom, and can be generated by energetic processing (photolysis or electron impact) of common interstellar ice constituents including H$_2$O, CO$_2$, O$_2$, and O$_3$ \citep{Stief1975, Slanger1971, Cosby1993, Lee1977, DeMore1966, Kedzierski2013}.  Of particular importance, Lyman-$\alpha$ photolysis of H$_2$O and CO$_2$ produces an O($^1$D) atom with 10\% and 100\% efficiency respectively.  As Ly-$\alpha$ is the dominant energy source in cold, shielded ISM regions, O($^1$D) atoms should be continuously produced in these ices.  Once formed, the O($^1$D) atom is meta-stable, with a gas-phase lifetime of $\sim$110s \citep{Sharpee2006}.  Measurements of the lifteime in astrophysically relevant matrices are lacking, but it is likely somewhat shorter: in neon and SF$_6$ matrices the O($^1$D) atom has a lifetime of 32s and $\sim$1s, respectively \citep{Fournier1982, Mohammed1990}.  Even so, a 1s lifetime is sufficient time for an O atom to scan $\sim$10$^5$ grain surface sites at a temperature of 15 K, assuming a diffusion barrier of 240 K \citep{Benderskii1995, Minissale2013} and a standard attempt frequency of 10$^{12}$ s$^{-1}$.  Theoretical and experimental work has demonstrated that O($^1$D) can diffuse within a matrix prior to reacting \citep{Ning2000, Tsuge2013}, and therefore collisional deactivation does not prohibit O($^1$D) diffusion-limited chemistry.

In \citet{Bergner2017}, we demonstrated that O($^1$D) atoms insert into the C-H bond of CH$_4$ to form CH$_3$OH or H$_2$CO with a negligible energy barrier.  Around 65\% of insertions led to CH$_3$OH, and the remainder to H$_2$CO.  Here we extend our study to the 2-carbon hydrocarbons C$_2$H$_6$, C$_2$H$_4$, and C$_2$H$_2$.  In this work, we aim to identify if the energetics of insertion proceed similarly in larger and unsaturated hydrocarbons, and to constrain the branching ratios of the various product channels.

The gas-phase chemistry of oxygen atoms with C$_2$ hydrocarbons is well-studied: typically the oxygen atom adds to or inserts into the hydrocarbon, followed by unimolecular dissociation of the excited product to form various radical species \citep[e.g.][]{Nunez2018}.  The chemistry in the solid state is less well understood, though a handful of studies have provided important clues.  Work by \citet{DeMore1969} explored O($^1$D) reactions with hydrocarbons in liquid argon matrices at 87 K, using O$_3$ photolysis as the O($^1$D) source.  The primary insertion product of O($^1$D) reactions with C$_2$H$_6$ was ethanol, while insertions into C$_2$H$_4$ yielded mainly acetaldehyde and ethylene oxide. Later, \citet{Parnis1993} confirmed the formation of ethanol by O($^1$D) insertion into C$_2$H$_6$ at 12 K, using UV-visible irradiation of N$_2$O:C$_2$H$_6$ in an argon matrix.   \citet{Schriver2007} similarly showed qualitatively that ethanol is formed from reactions of oxygen atoms with C$_2$H$_6$ at 11 K, using O$_3$ or CO$_2$ photolysis as the oxygen source.  \citet{Hawkins1983} used O$_3$ irradiation by an Hg arc lamp to study O + C$_2$H$_4$ in argon matrices at 20 K, and identified products including acetaldehyde, ethylene oxide, and vinyl alcohol.  More recently, \citet{Bennett2005b} radiolysed C$_2$H$_4$ + CO$_2$ ice with 5 keV electrons at 11 K to explore the formation mechanisms of acetaldehyde, ethylene oxide, and vinyl alcohol.  CO$_2$ $\rightarrow$ CO + O was the presumed chemical driver, but the electronic state of oxygen responsible for the chemistry was not isolated.   \citet{Ward2011} isolated the reactions of the ground-state O($^3$P) atoms with C$_2$H$_4$ at temperatures from 12--90 K, finding ethylene oxide and acetaldehyde as the major products.  \citet{Haller1962} studied O($^3$P) reactions with C$_2$H$_2$ to form ketene in argon matrix conditions at 20K.  \citet{Hudson2013} recently showed that ketene can be formed in radiolysed interstellar ice analogs containing C$_2$H$_2$ and oxygen-bearing molecules.

These studies together suggest that hydrocarbon reactions with oxygen may be important in ISM ices, but key questions remain:  (i) What are the kinetics and energetics of organic product formation?  (ii) What are the branching ratios of product formation?  (iii) Do hydrocarbon dissociation products play a role, or are the observed products mainly formed through a direct O + hydrocarbon mechanism? (iv) What is the role of the excited vs. the ground state oxygen atom? (v) How does the reaction mechanism differ for single, double, and triple bonded hydrocarbons?

In this work we perform a systematic study of how photo-produced oxygen atoms react with ethane, ethylene, and acetylene to address these unresolved issues, and to evaluate the astrochemical importance of oxygen atom reactions with C$_2$ hydrocarbons.

\section{Methods}
\subsection{Experimental setup}
Experiments are performed using an ultra-high vacuum setup with a base pressure of $\sim5 \times$ 10$^{-10}$ Torr, described in detail in \citet{Lauck2015}.  A CsI substrate window is cooled by a closed-cycle He cryostat to 9 K.  A LakeShore 335 temperature controller is used to monitor and control the substrate temperature, with an estimated accuracy of 2 K and a relative accuracy of 0.1 K.  A Pfeiffer QMG 220M1 quadrupole mass spectrometer (QMS) is used to monitor gas-phase species in the chamber.  Ices are grown on the substrate by introducing gases through a dosing pipe.  Gases are mixed in a differentially pumped gas line with a base pressure of $\sim$10$^{-4}$ Torr.  We use the following gases in our experiments: CO$_2$ (Aldrich, 99.9 atom \% $^{12}$C), $^{13}$CO$_2$ (Aldrich, 99 atom \% $^{13}$C, $<$3 atom \% $^{18}$O), C$^{18}$O$_2$ (Aldrich, 97 atom \% $^{18}$O), $^{13}$CO (Aldrich, $\leq$5 atom \% $^{18}$O, 99 atom \% $^{13}$C), C$^{18}$O (95 atom \% $^{18}$O), C$_2$H$_6$ (Aldrich, 99.99\%), C$_2$H$_4$ (Aldrich, $\geq$99.5\%), and C$_2$H$_2$ (99.6\% Matheson Tri-gas, dissolved in acetone\footnote{Acetone was not removed prior to dosing, and was detected at very low levels ($<$1\%) with respect to C$_2$H$_2$ based on the IR feature at 1710 cm$^{-1}$ \citep{Hudson2018}.}).

\subsection{Ice column densities}
\label{sec:columns}
A Bruker Vertex 70v Fourier transform infrared spectrometer is used in transmission mode to measure the column density of infrared-active molecules in the ice.  For each IR spectrum, 128 interferograms were collected, averaged, and transformed.  Molecule column densities $N_i$ are calculated as:
\begin{equation}
N_i = \frac{2.3 \int Abs (\tilde{\nu}) d\tilde{\nu}}{A_i},
\end{equation}
where $\int Abs(\tilde{\nu}) d\tilde{\nu}$ is the area of the IR band in absorbance and $A_i$ is the band strength in optical depth. 

The band strengths and peak centers used for this analysis are listed in Table \ref{tab:irdat}.  For consistency, we adopt the band strength of the major isotopologue for all minor isotopologues, since measurements are not available for all minor species.  We note that in the case of acetaldehyde, the only experimentally derived solid-state band strength is for a $\sim$20:1 H$_2$O:CH$_3$CHO mixture measured with reflection spectroscopy \citep{Moore1998}.  Without a robust way to convert to a transmission band strength, we adopt the theoretical value from \citet{Bennett2005a}, which differs by $\sim$30\% from the measured reflection value.  Similarly, no solid-state band strength is available for ethylene oxide and so we adopt the gas-phase value, which agrees within 10\% with the theoretical value presented in \citet{Bennett2005b}.  For both acetaldehyde and ethylene oxide, the lack of a solid-state transmission band strength introduces large uncertainties into the calculation of column densities.  For all other molecules the formal band strength uncertainties are $\sim$20\%.  However, the true band strength uncertainties may be higher since ice composition can impact the band intensity, and the values listed in Table \ref{tab:irdat} were derived for pure ices.

\begin{deluxetable}{llll} 
	\tabletypesize{\footnotesize}
	\tablecaption{IR peak positions and band strengths \label{tab:irdat}}
	\tablecolumns{4} 
	\tablewidth{\textwidth} 
	\tablehead{
		\colhead{Molecule}                            &
		\colhead{Line center}                         &
		\colhead{Band strength}                     &
		\colhead{Ref.}                                     \\
		\colhead{}                                           & 
		\colhead{(cm$^{-1}$)}                           & 
		\colhead{(cm molecule$^{-1}$)}           &
		\colhead{} }
\startdata
C$_2$H$_6$     & 1462 & 3.76 $\times$ 10$^{-18}$ & 1 \\
                          & 2880 & 3.81 $\times$ 10$^{-18}$ & 1 \\ 
C$_2$H$_4$     & 949   & 1.28 $\times$ 10$^{-17}$ & 1 \\
C$_2$H$_2$     & 742   & 2.42 $\times$ 10$^{-17}$   & 2  \\
CH$_4$             & 1302  & 8.0 $\times$ 10$^{-18}$  & 3   \\
CO$_2$ & 2342 & 7.6 $\times$ 10$^{-17}$   &  3 \\
$^{13}$CO$_2$ & 2283 &  & \\
C$^{18}$O$_2$ & 2306 & & \\
CO         & 2139 & 1.12 $\times$ 10$^{-17}$   & 3  \\
$^{13}$CO & 2092 & & \\
C$^{18}$O & 2088 & & \\
C$_2$H$_5$OH & 1050 & 1.41 $\times$ 10$^{-17}$ & 4 \\
CH$_3$CHO      & 1351  & 4.5 $\times$ 10$^{-18}$ $^{a}$  & 5 \\
c-C$_2$H$_4$O & 872   & 1.2 $\times$ 10$^{-17}$ $^{b}$ & 6 \\
CH$_2$CO      & 2133 & 1.2 $\times$ 10$^{-16}$ &  7 \\
CH$_2$C$^{18}$O & 2107 & &  \\
\enddata
\tablecomments{Isotopologues are assumed to have the same band strength.  References: [1] \citet{Hudson2014}, [2] \citet{Hudson2014a}, [3] \citet{Bouilloud2015}, [4] \citet{Hudson2017}, [5] \citet{Bennett2005a}, [6] \citet{Nakanaga1981}, [7] \citet{Berg1991}}
\tablenotetext{\rm{a}}{Theoretical value}
\tablenotetext{\rm{b}}{Gas-phase value}
\end{deluxetable}

\subsection{UV Lamp}
\label{sec:lamp}
An H$_2$:D$_2$ lamp (Hamamatsu L11798) is used to irradiate ice samples.  We use a NIST-calibrated AXUV-100G photodiode to measure the photon flux at the sample holder; the flux uncertainties are $\sim$5\% for the wavelengths used in this work.  Since VUV photons ($<$200nm) are responsible for the observed chemistry, we estimate the VUV component of the lamp flux as follows.  We first measure the flux of photons with wavelengths above 200nm by filtering the lamp with a MgO window.  This $>$200nm flux is then subtracted from the total flux to obtain the flux of VUV photons at the sample.  Because of this subtraction, the true uncertainty in flux measurement is likely closer to 10\%.  Experiments were performed with no lamp filter in place.  The VUV flux measurements for each experiment are listed in Table \ref{tab:expsumm}.

The radiation dose absorbed by the sample is estimated by solving for the absorbed photon flux: 

\begin{equation}
I_{a,i}(\lambda) = I_{o}(\lambda)(1-e^{-N_i\sigma_i(\lambda)}),
\end{equation}

\noindent where $I_{o}(\lambda)$ is the incident intensity, $I_{a,i}(\lambda)$ is the intensity absorbed by species $i$, $N_i$ is the column density of species $i$, and $\sigma_i$ is the wavelength-dependent UV absorption cross section.  We use the initial column density of each species following dosing to determine the flux and in turn energy absorbed in the wavelength range from 110 -- 180nm.  Solid-phase CO and CO$_2$ VUV absorption cross-sections are taken from \citet{CruzDiaz2014a} and \citet{CruzDiaz2014b}, respectively.  We use gas-phase cross-sections for the hydrocarbons since solid-phase cross-sections are not available.  C$_2$H$_6$, C$_2$H$_4$, and C$_2$H$_2$ cross-sections are taken from \citet{Au1993}, \citet{Lu2004}, and \citet{Cooper1995}, via the MPI-Mainz Spectral Atlas \citep{Keller2013}.  The resulting UV dose rates are listed in Table \ref{tab:expsumm}.

\subsection{Experimental scheme}
Our aim is to study the interactions of O($^{1}$D) atoms with the hydrocarbons C$_2$H$_6$, C$_2$H$_4$, and C$_2$H$_2$.  We use CO$_2$ as our O($^1$D) atom donor since the dissociation products for wavelengths 120 -- 167.2 nm is CO + O($^1$D) with a quantum yield of unity, and the CO$_2$ absorption cross-section at longer wavelengths is negligible \citep{Okabe1978}.  All O atoms generated by CO$_2$ irradiation with the H$_2$:D$_2$ lamp are therefore O($^1$D) atoms.  O($^3$P) atoms will form through deexcitation of O($^1$D) atoms that do not react on short timescales.  

For C$_2$H$_6$ and C$_2$H$_4$ experiments, we use $^{13}$CO$_2$ to enable identification of any chemistry involving the side-product $^{13}$CO based on the $^{13}$C label.  C$^{18}$O$_2$ was used for C$_2$H$_2$ experiments to enable the IR features of ketene and CO to be resolved.

The solid-state UV absorption cross sections of the C$_2$ hydrocarbons have not been measured, however in the gas phase they absorb broadly over the spectrum of the H$_2$:D$_2$ lamp \citep{Au1993, Lu2004, Cooper1995}.  Therefore, hydrocarbon dissociation side-products will be produced when generating O($^1$D) in situ.  We perform control experiments of irradiated hydrocarbon:CO mixtures in order to constrain the reactivity of all products excluding oxygen atoms, thereby isolating the chemistry driven by oxygen atoms in the ice.

For each experiment, a pure or mixed ice was deposited on the substrate at 9 K.  Samples were irradiated for 2--3 hours at a set temperature, with IR scans taken every 3 minutes.  Following irradiation the sample was heated at a rate of 5 K minute$^{-1}$ to 200 K, with IR scans every 2 minutes.  All experimental details are summarized in Table \ref{tab:expsumm}.  Mixing ratios are derived based on the IR-measured column density of each ice component.  All IR spectra obtained during our experiments are available on Zenodo \citep{Zenodo}.

\begin{deluxetable*}{lcccccc} 
	\tabletypesize{\footnotesize}
	\tablecaption{Experiment summary \label{tab:expsumm}}
	\tablecolumns{7} 
	\tablewidth{\textwidth} 
	\tablehead{
		\colhead{Ice composition}                   &
		\colhead{Experiment}                         &
		\colhead{Irradiation temperature}        &
		\colhead{Ratio}                                    &
		\colhead{Total column density}            &
		\colhead{Incident UV flux}                                &
		\colhead{Absorbed UV dose$^a$}              \\
		\colhead{}                                           & 
		\colhead{}                                          & 
		\colhead{(K)}                                      & 
		\colhead{}                                          &
		\colhead{(ML)}                                   &
		\colhead{(10$^{13}$ ph cm$^{-2}$ s$^{-1}$)} &
		\colhead{(10$^{13}$ eV cm$^{-2}$ s$^{-1}$)}}
\startdata
C$_2$H$_6$ & 1 & 9 & - &  46 & 7.9 & 12 \\
C$_2$H$_4$ & 2 & 9 & - &  37 & 9.2 & 37 \\
C$_2$H$_2$ & 3 & 9 & - &  35 & 5.7 & 22 \\
$^{13}$CO$_2$ & 4 & 9 & - & 121 & 6.3 & 5 \\
\hline
C$_2$H$_6$:$^{13}$CO & 5 & 9 & 0.5:1 & 178 & 5.9 & 18 \\
C$_2$H$_4$:$^{13}$CO & 6 & 9 & 0.7:1 & 188 & 5.3 & 46 \\
C$_2$H$_2$:C$^{18}$O & 7 & 9 & 0.5:1 & 185 & 4.6 & 33 \\
\hline
C$_2$H$_6$:$^{13}$CO$_2$ & 8 & 9 & 0.5:1 & 201 & 6.9 & 18 \\
  & 9 & 9 & 0.8:1 & 213 & 7.1  & 21 \\
  & 10 & 14 & 0.5:1 & 176 & 6.4  & 16 \\
  & 11 & 19 & 0.5:1 & 169 & 6.5  & 15 \\
  & 12 & 24 & 0.6:1 & 151 & 8.6  & 18 \\
  & 13 & 24 & 0.6:1 & 166 & 5.9  & 14 \\
  & 14 & 40 & 0.5:1 & 192 & 4.7  & 16 \\
C$_2$H$_6$:C$^{18}$O$_2$ & 15 & 9 & 0.5:1 & 171 & 4.3 & 14 \\
\hline
C$_2$H$_4$:$^{13}$CO$_2$ & 16 & 9 & 0.6:1 & 169 & 9.1 & 54 \\
  & 17 & 9 & 0.7:1 & 181 & 7.2  & 48 \\
  & 18 & 14 & 0.7:1 & 176 & 6.7  & 46 \\
  & 19 & 19 & 0.7:1 & 172 & 7.1  & 46 \\
  & 20 & 24 & 0.7:1 & 177 & 6.6  & 46 \\
  & 21 & 40 & 0.7:1 & 197 & 5.9  & 45 \\
  & 22 & 52 & 0.7:1 & 195 & 5.0  & 42 \\
  & 23 & 55 & 0.7:1 & 205 & 5.1  & 44 \\
\hline
C$_2$H$_2$:C$^{18}$O$_2$ & 24 & 9 & 0.5:1 & 227 & 4.9 & 38 \\
  & 25 & 14 & 0.5:1 & 243 & 5.1  & 40 \\
  & 26 & 19 & 0.5:1 & 239 & 4.5  & 37 \\
  & 27 & 24 & 0.5:1 & 230 & 5.6  & 41 \\
  & 28 & 40 & 0.5:1 & 229 & 4.7  & 37 \\
  & 29 & 52 & 0.5:1 & 242 & 4.4  & 37 \\
\enddata
\tablenotetext{\rm{a}}{Calculated for the initial ice composition and wavelengths from 110 -- 180nm.}
\end{deluxetable*}

\section{Results}
\subsection{Irradiation products}
\label{sec:irr_pr}

\begin{figure}
	\includegraphics[width=0.9\linewidth]{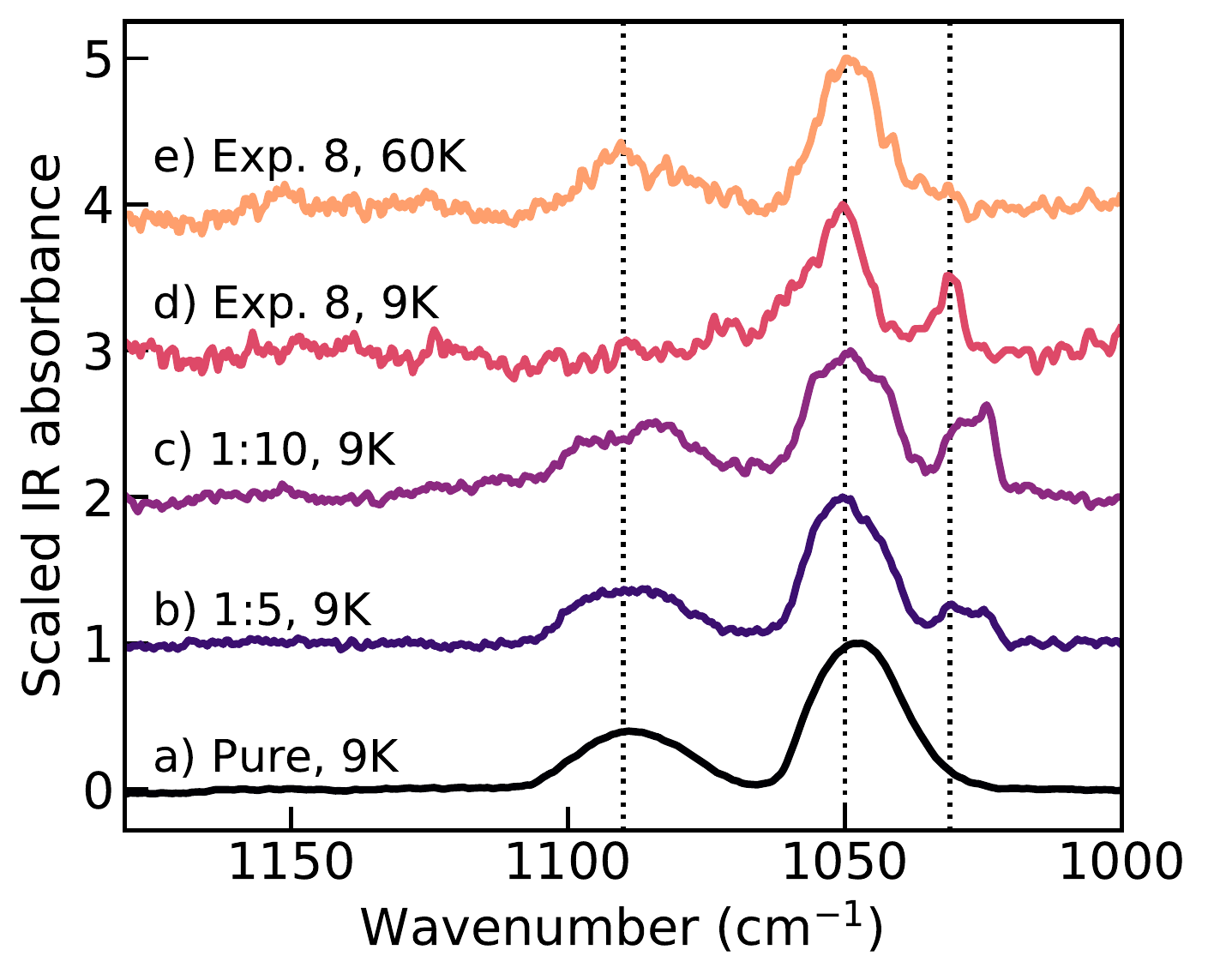}
	\caption{IR features of ethanol. (a) Pure ethanol at 9 K, (b) a 1:5 ethanol:ethane mixture at 9 K, and (c) a 1:10 ethanol:ethane mixture at 9 K.  Spectra (e) and (d) show an irradiated $^{13}$CO$_2$:C$_2$H$_6$ ice mixture at 9 K and upon heating to 60 K.  Dotted vertical lines mark ethanol features at 1090 cm$^{-1}$ (only pure ethanol), 1050 cm$^{-1}$ (pure and diluted in ethane), and 1030 cm$^{-1}$ (only diluted in ethane).}
\label{fig:ethanol}
\end{figure}

Reaction products are identified by their IR features.  In most cases, product assignment is straightforward as the peak positions agree well with literature values.  We note that when ethanol is mixed with C$_2$H$_6$, the 1090 cm$^{-1}$ feature seen in the pure spectrum is diminished and a new feature around 1030 cm$^{-1}$ appears (Figure \ref{fig:ethanol}).  This explains why the ethanol feature in irradiated C$_2$H$_6$:$^{13}$CO$_2$ mixtures does not match the pure spectrum at low temperatures, but above the C$_2$H$_6$ desorption temperature it converges to its pure spectrum. 

\begin{figure*}
	\includegraphics[width=\linewidth]{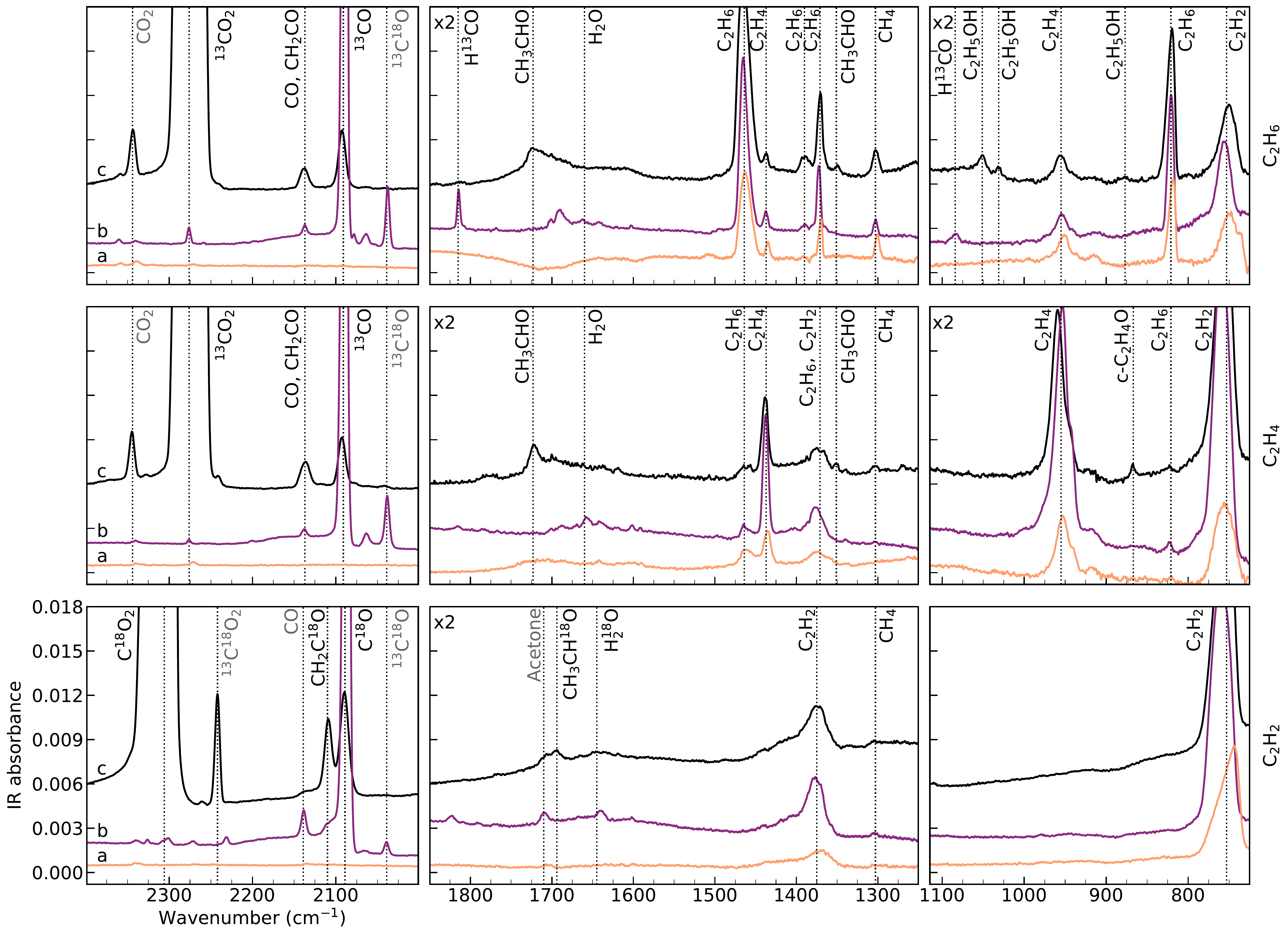}
	\caption{Post-irradiation IR spectra for C$_2$H$_6$ (top), C$_2$H$_4$ (middle), and C$_2$H$_2$ (bottom) experiments.  For each hydrocarbon, irradiations of pure ices (a), mixtures with CO (b), and mixtures with CO$_2$ (c) are shown.  C$_2$H$_6$ and C$_2$H$_4$ are mixed with $^{13}$CO and $^{13}$CO$_2$ and C$_2$H$_2$ is mixed with C$^{18}$O and C$^{18}$O$_2$.  Panels marked with ``$\times$2'' have been scaled for clarity.  Isotopic and other contaminants are indicated with gray text.}
\label{fig:ir_diff}
\end{figure*}

Figure \ref{fig:ir_diff} shows the IR spectra following $\sim$2h irradiation at 9 K for C$_2$H$_6$, C$_2$H$_4$, and C$_2$H$_2$, pure and mixed with CO and CO$_2$.  Additionally, the pre- and post-irradiation spectra for the hydrocarbon:CO$_2$ experiments are shown in the Appendix.  The products observed for each mixture are as follows.  

\textit{C$_2$H$_6$}  
Irradiations of pure C$_2$H$_6$ yield the hydrocarbons CH$_4$, C$_2$H$_4$, and C$_2$H$_2$.  When C$_2$H$_6$:$^{13}$CO mixtures are irradiated, we also observe the formation of the H$^{13}$CO radical.  For C$_2$H$_6$:$^{13}$CO$_2$ mixtures, in which oxygen atoms are generated, the additional organics ethanol (C$_2$H$_5$OH) and acetaldehyde (CH$_3$CHO) are formed, as well as $^{13}$CO, $^{12}$CO$_2$ and $^{12}$CO.  Ketene (CH$_2$CO), whose strongest IR feature overlaps with $^{12}$CO, is detected upon CO sublimation, and confirmed in a C$_2$H$_6$:C$^{18}$O$_2$ irradiation.

\textit{C$_2$H$_4$}
For both pure C$_2$H$_4$ and $^{13}$CO:C$_2$H$_4$ irradiations, the main products are C$_2$H$_6$ and C$_2$H$_2$.  Irradiations of C$_2$H$_4$:$^{13}$CO$_2$ additionally produce the organics ethylene oxide (c-C$_2$H$_4$O) and acetaldehyde.  As in the case of C$_2$H$_6$, we also observe $^{13}$CO, $^{12}$CO, and $^{12}$CO$_2$.  Ketene is also detected following CO sublimation.

\textit{C$_2$H$_2$}  Pure C$_2$H$_2$ irradiations do not result in any obvious features in the IR, although C$_2$H$_2$ consumption is apparent.  C$_2$H$_2$:C$^{18}$O irradiations result in the production of several minor CO and CO$_2$ isotopologue features.  For C$_2$H$_2$:C$^{18}$O$_2$ mixtures, we observe the formation of CH$_2$C$^{18}$O and C$^{18}$O.

\subsection{Kinetic modeling}
\label{sec:model}
\begin{figure}
	\includegraphics[width=\linewidth]{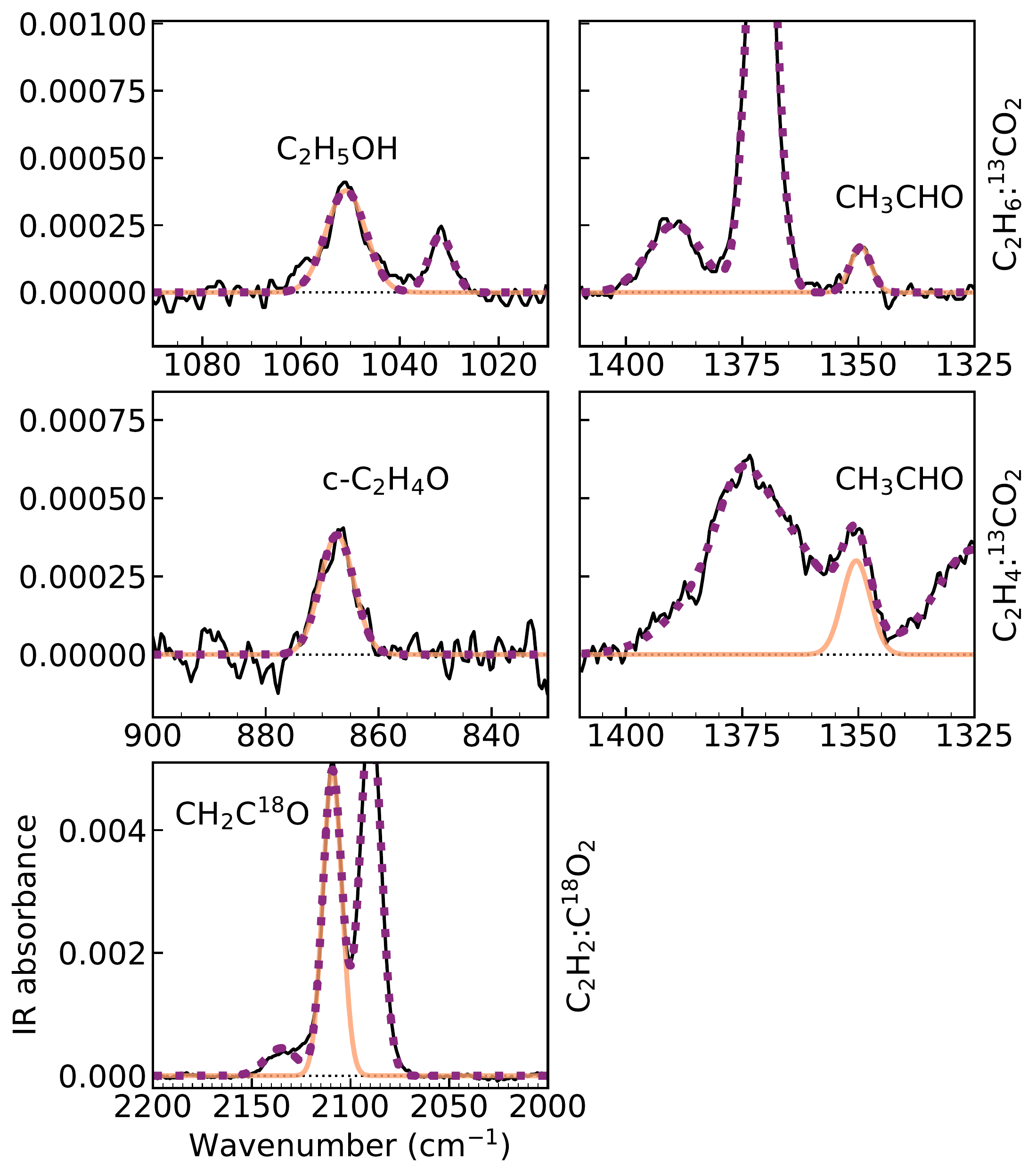}
	\caption{Example fits to the main organic products of C$_2$H$_6$:$^{13}$CO$_2$ irradiations (top), C$_2$H$_4$:$^{13}$CO$_2$ irradiations (middle), and C$_2$H$_2$:C$^{18}$O$_2$ irradiations (bottom).  The fit to the target peak is shown in orange, and the total fit including any neighboring lines is shown in purple dashes.}
\label{fig:irfit}
\end{figure}

Infrared spectra are used to derive growth curves for product molecules over the course of each irradiation.  The main infrared peaks used to quantify each species are listed in Table \ref{tab:irdat}.  For the organic products arising from oxygen atom chemistry (ethanol, acetaldehyde, ethylene oxide, and ketene), we fit a Gaussian to the target IR feature along with any overlapping neighbor features and a linear baseline; example fits are shown in Figure \ref{fig:irfit}.  

Growth curves for the main organic products are well-described by a simple steady-state kinetic model:
\begin{equation}
\label{eq:gc}
[\mathrm{product}](t) = N_{ss}(1 - e^{-k_{r}\phi}),
\end{equation}
where $k_{r}$ is the reaction rate constant and $N_{ss}$ is the steady-state abundance.  We fit each growth curve using photon fluence $\phi$ instead of time in order to account for differences in the measured lamp flux for a given experiment.  Growth curves for C$_2$H$_6$ and C$_2$H$_4$ experiments are fit in the range of 0 to 4.8 $\times$ 10$^{17}$ photons cm$^{-2}$.  For C$_2$H$_2$ experiments, growth curves are fit from 0 to 2.4 $\times$ 10$^{17}$ photons cm$^{-2}$ since the curves are less well-described by first-order kinetics after this.

\begin{figure}
	\includegraphics[width=\linewidth]{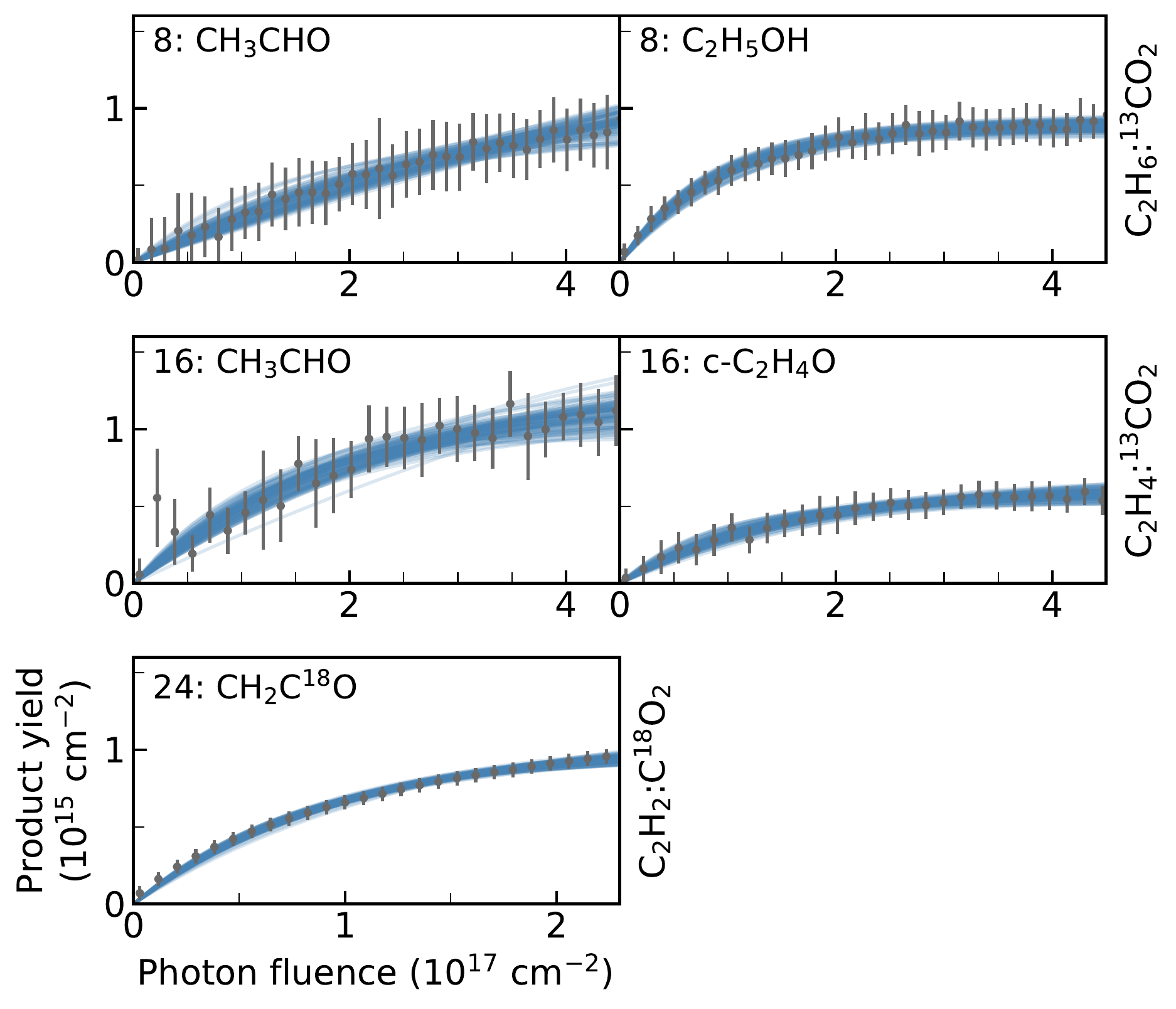}
	\caption{Example growth curves with kinetic model fits for the main organic products of experiments 8, 16, and 24.  Experimental data is shown as grey points, and blue lines show fits drawn from the posterior probability distribution.}
\label{fig:gc_ex}
\end{figure}

We use the Markov-chain monte carlo code {\fontfamily{qcr}\selectfont emcee} \citep{Foreman-Mackey2013} to fit each growth curve.  With this method we can evaluate the degeneracy of our fit parameters and better constrain their uncertainties.   Figure \ref{fig:gc_ex} shows example growth curves and best-fit kinetic models for each experimental series; all additional growth curves are shown in the Appendix, along with example corner plots for each product.  The best-fit parameters and uncertainties derived from growth curve fitting are also listed in the Appendix.

We also estimate the photolysis timescales for pure hydrocarbon and CO ices by fitting the loss curves from Experiments 1--4 as exponential decays.  We find that the timescales for photolysis of C$_2$H$_6$, C$_2$H$_4$, C$_2$H$_2$, and $^{13}$CO molecules are $\sim$300 minutes, 70 minutes, 170 minutes, and 900 minutes.  Thus, at early times in our experiments it is unlikely that a given molecule will undergo multiple dissociations.

\subsection{Temperature dependence}
\label{sec:arrh}
\begin{figure}
	\includegraphics[width=\linewidth]{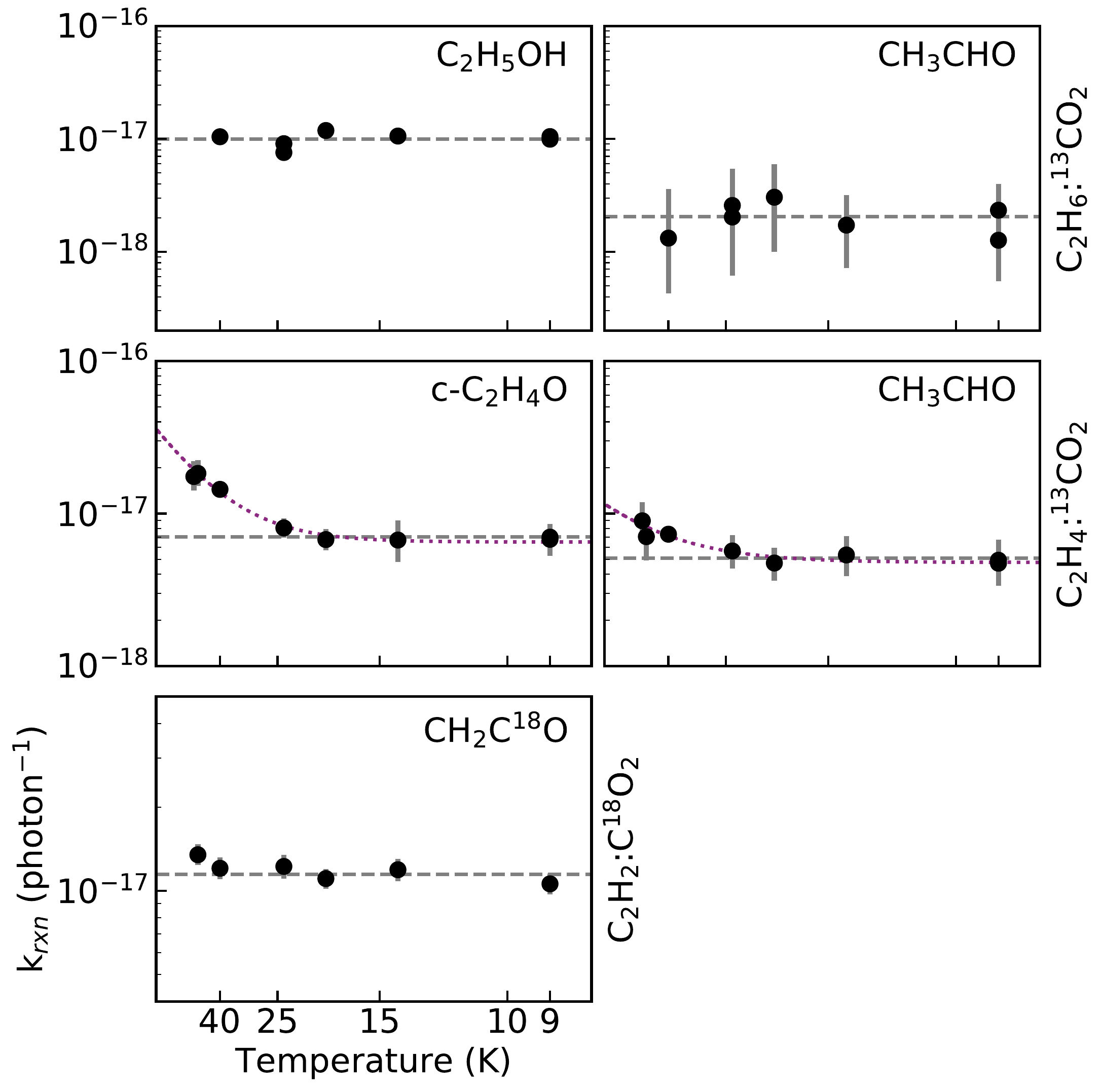}
	\caption{Reaction rate constants derived from growth curve fits plotted against the temperature of ice irradiation for C$_2$H$_6$:$^{13}$CO$_2$ (top), C$_2$H$_4$:$^{13}$CO$_2$ (middle), and C$_2$H$_2$:C$^{18}$O$_2$ (bottom).  Dashed grey lines show the mean rate constant for all temperatures, excluding points above 25 K for C$_2$H$_4$ products.  Dotted purple lines in the C$_2$H$_4$ row show a two-component fit consisting of a flat and Arrhenius-like contribution.}
\label{fig:arrh}
\end{figure}

Figure \ref{fig:arrh} shows Arrhenius plots for the reaction rate constants derived from growth curve fitting.  For the C$_2$H$_6$ experiments, the C$_2$H$_5$OH formation rate constants show no temperature dependence from 9 K -- 40 K.  CH$_3$CHO rate constants likewise appear temperature independent, although with higher scatter due to uncertainties in fitting the acetaldehyde spectral feature.  Similarly, ketene formation rate constants in the C$_2$H$_2$ experiments also show no obvious temperature dependence from 9 K -- 52 K.  For the C$_2$H$_4$ experiments, we see no temperature dependence to the ethylene oxide or acetaldehyde rate constants at low temperatures (9 K -- 25 K), but an increasing rate at higher temperatures.  We fit this temperature dependence using a two-component model consisting of a flat component $k_0$ and a barriered component:
\begin{equation}
k_{r} = Ae^{-E_a/T} + k_0,
\end{equation}
and find barriers of 84 $\pm$ 13 K for ethylene oxide and 63 $\pm$ 25 K for acetaldehyde.

\subsection{Steady-state carbon budget}
\label{sec:yields}

\begin{figure*}
\centering
	\includegraphics[width=0.9\linewidth]{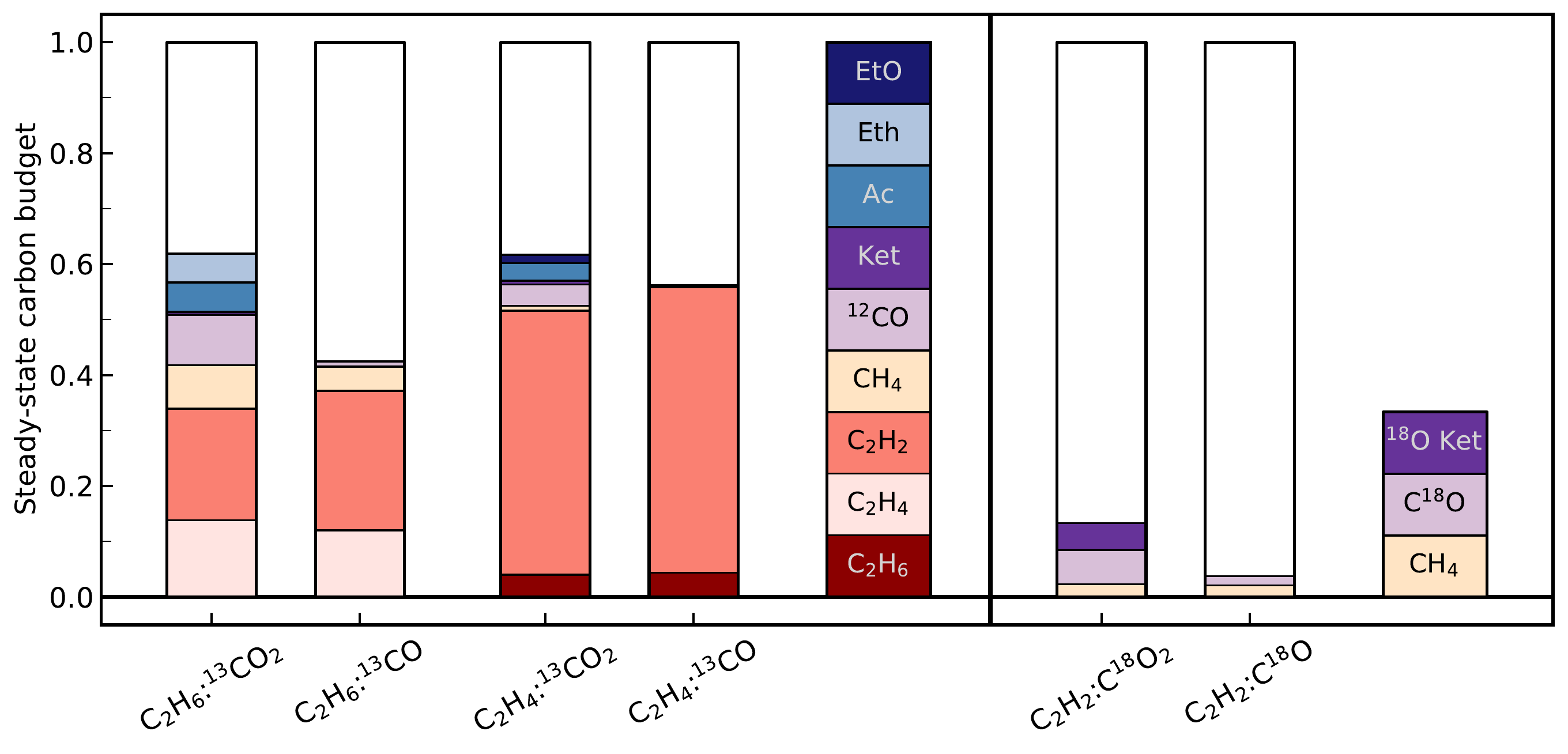}
	\caption{Steady-state carbon atom budget for 9 K irradiations.  The fractional yield for each experiment shows the number of carbon atoms in each product as a fraction of the number of carbon atoms lost from destruction of the reactant hydrocarbon. The far right bars show the product color key.  EtO = ethylene oxide, Eth = ethanol, Ac = acetaldehyde, and Ket = ketene.}
\label{fig:budget}
\end{figure*}

Figure \ref{fig:budget} shows the fractional yield of carbon-bearing products relative to the parent hydrocarbon consumption, i.e. products with a a single carbon count for half of a consumed C$_2$ hydrocarbon molecule.  To determine steady-state product yields, we use the IR spectrum taken at the final time point of the growth curves ($\sim$4.8 $\times$ 10$^{17}$ ph cm$^{-2}$ for C$_2$H$_6$ and C$_2$H$_4$ experiments and 2.4$\times$10$^{17}$ph cm$^{-2}$ for C$_2$H$_2$ experiments).  Column densities for the organic products are calculated as described in Section \ref{sec:model}, and for the remaining products by integrating the IR features listed in Table \ref{tab:irdat}.

We attempt to count only carbon that originates from the hydrocarbons and not the CO$_2$; this is straightforward for C$_2$H$_6$:$^{13}$CO$_2$ and C$_2$H$_4$:$^{13}$CO$_2$ experiments because of the isotopic label, but somewhat more challenging for the C$_2$H$_2$:C$^{18}$O$_2$ experiments.  In this case, C$^{18}$O could be formed from C$^{18}$O$_2$ dissociation, or from carbon originating in C$_2$H$_2$.  We estimate the amount of C$^{18}$O formed from C$_2$H$_2$ carbon by subtracting the C$^{18}$O$_2$ loss from the total C$^{18}$O yield, since each CO$_2$ dissociation produces a CO molecule, and CO should not undergo other chemistry at a significant rate.  

For C$_2$H$_6$ and C$_2$H$_4$ experiments, the ketene and $^{12}$CO IR features are overlapped.  We therefore estimate the ketene yield from the 2133 cm$^{-1}$ feature at 70 K following CO desorption.  Some CO may be trapped in the ice still at this temperature; we therefore use the percentage of trapped $^{13}$CO at 70 K to estimate the amount of trapped $^{12}$CO, and subtract the corresponding IR area from the ketene peak area.  

From this analysis, we derive the following steady-state product ratios for hydrocarbon:CO$_2$ experiments.  For C$_2$H$_6$:$^{13}$CO$_2$, we find 0.14 C$_2$H$_4$ : 0.20 C$_2$H$_2$ : 0.04 CH$_4$ : 0.09 CO : 0.01 CH$_2$CO : 0.05 CH$_3$CHO : 0.05 C$_2$H$_5$OH.  For C$_2$H$_4$:$^{13}$CO$_2$, we find 0.03 C$_2$H$_6$ : 0.47 C$_2$H$_2$ : 0.04 CO :  0.01 CH$_2$CO : 0.03 CH$_3$CHO : 0.01 c-C$_2$H$_4$O.  Lastly, for C$_2$H$_2$:C$^{18}$O$_2$ we find 0.01 CH$_4$ : 0.07 C$^{18}$O : 0.04 CH$_2$C$^{18}$O.  In all cases, the yield of hydrocarbon products is similar between the CO (control) experiments and the CO$_2$ (oxygen atom) experiments.  This indicates that the formation of hydrocarbons in these experiments is largely side chemistry and is not related to oxygen atom chemistry.  The ratios of organic products will be discussed in detail in Section \ref{sec:br_ratio}.

We note that the sum of products does not fully account for the consumption of reactants.  One possible origin of the missing carbon is band strength uncertainties: in addition to the uncertainties discussed in Section \ref{sec:columns}, \citet{Terwisscha2017} have shown that the CH$_3$CHO and C$_2$H$_5$OH band strengths are quite sensitive to the ice environment, so it is possible that we are under-estimating the yields of the organic products.  Photodesorption may also contribute to the observed carbon deficit, though without measured photodesorption cross-sections for the C$_2$ hydrocarbons it is difficult to estimate its importance.  Assuming the CH$_4$ photodesorption rate \citep{Dupuy2017} and given the H$_2$:D$_2$ lamp spectrum, we estimate that at most $\sim$1\% of the consumed carbon could be photodesorbed.  However, CH$_4$ absorbs more weakly and in a much narrower wavelength range than the C$_2$ hydrocarbons, and so it is possible that the photodesorption is significant in our experiments.  Partitioning of carbon into radical fragments or longer carbon chains that are not readily detected by IR spectroscopy could also account for some of the missing carbon.

\subsection{Primary products \& reaction network}
\label{sec:pr_pr}
\begin{figure*}
	\includegraphics[width=\linewidth]{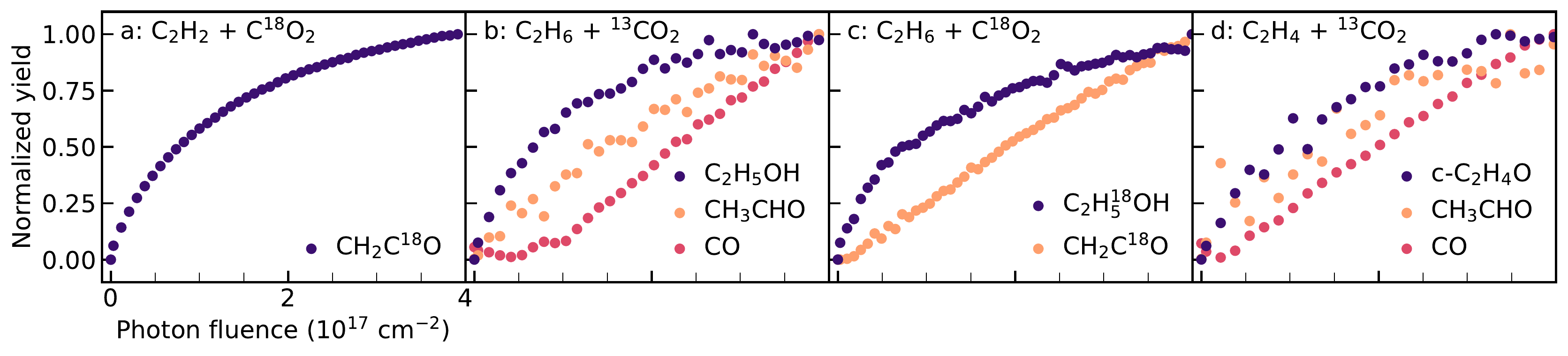}
	\caption{Normalized growth curves for the products of hydrocarbon + O reactions.}
\label{fig:vsgc}
\end{figure*}

We now discuss the products resulting from O + hydrocarbon chemistry, and use growth curve shapes (Figure \ref{fig:vsgc}) to identify primary and secondary products.  Again, based on Figure \ref{fig:budget} we do not consider the formation of hydrocarbon products to be related to oxygen atom chemistry.

The products of C$_2$H$_2$ + O are ketene and CO.  Ketene's growth curve shows a rapid and early formation (Figure \ref{fig:vsgc}a), indicating that it is a first-generation product of O + C$_2$H$_2$.  We cannot isolate a C$^{18}$O growth curve since C$^{18}$O is produced in large quantities from C$^{18}$O$_2$ photolysis.  It is therefore unclear whether CO is a primary or secondary product of O + C$_2$H$_2$.  CO could be formed partly from ketene photodissociation, however its high abundance (twice that of ketene) is difficult to explain purely from a ketene destruction pathway.  Oxygen atom reactions with hydrocarbon fragments could also contribute to the CO yield.  It is also possible that CO + CH$_2$ is a direct exit channel of O + C$_2$H$_2$ as seen in the gas phase \citep[e.g.][]{Nunez2018}, however we are not able to confirm that this primary channel exists without isolating a CO growth curve.

The products of C$_2$H$_6$ + O are ethanol, acetaldehyde, ketene, and CO.  Ethanol and acetaldehyde appear early in the experiment ($<$0.1$\times$10$^{17}$ph cm$^{-2}$; Figure \ref{fig:vsgc}b), indicative that they have a first-generation formation channel.  The CO growth curve shows a delay in formation compared to the first-generation products, and so CO is likely formed from downstream chemistry such as photolysis of the primary organic products, or O atom reactions with hydrocarbon fragments.   We are able to resolve the ketene growth curve from CO in a C$_2$H$_6$:C$^{18}$O$_2$ experiment, and find that ketene formation lags ethanol formation (Figure \ref{fig:vsgc}c), indicating that it is not a primary product of O + C$_2$H$_6$.  Since C$_2$H$_2$ is produced in abundance as a side-product in these experiments, it is likely that some of the ketene formation is due to secondary C$_2$H$_2$ + O reactions.  Ketene could also be formed by photolysis of the primary organic products.

The products of C$_2$H$_4$ + O are ethylene oxide, acetaldehyde, ketene, and CO.  The early formation of ethylene oxide and acetaldehyde indicates that they are primary products (Figure \ref{fig:vsgc}d).  Similar to C$_2$H$_6$, CO and ketene are likely second-generation products of C$_2$H$_4$ + O based on the delay in their formation.

\begin{figure}
	\includegraphics[width=\linewidth]{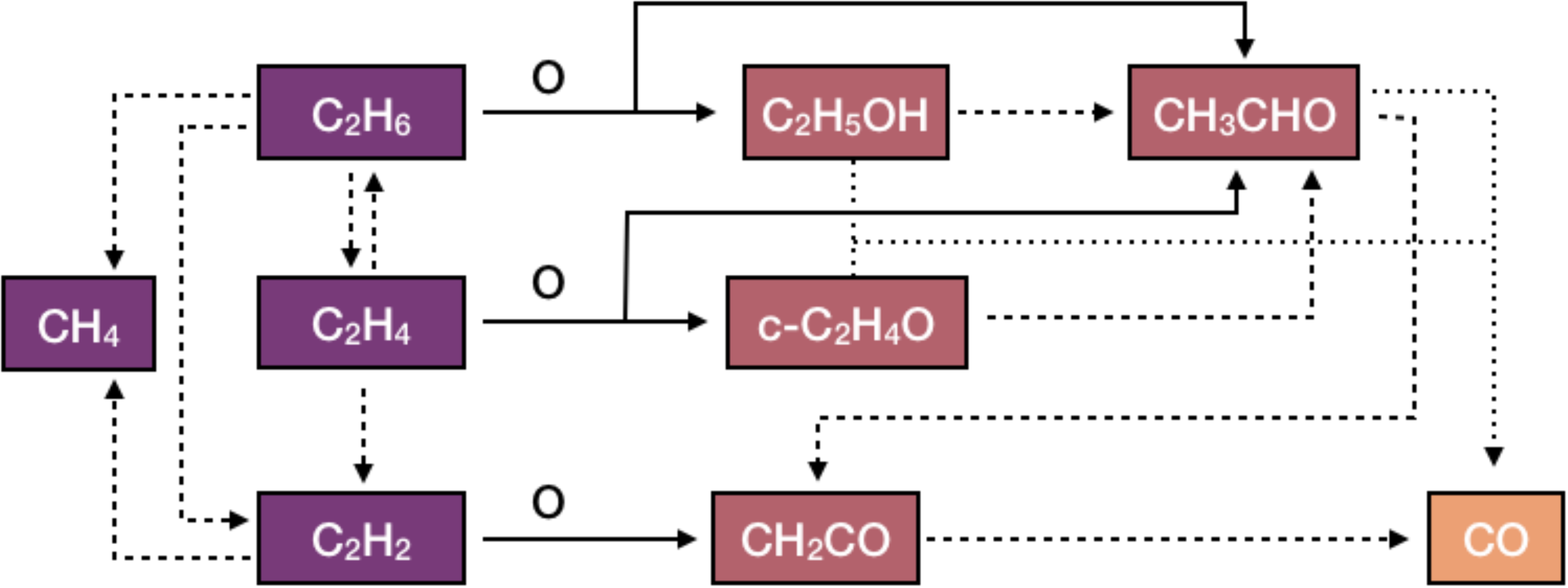}
	\caption{Reaction network for oxygen atom reactions with C$_2$H$_6$, C$_2$H$_4$, and C$_2$H$_2$.  Primary reaction pathways for hydrocarbons + oxygen atoms are shown as solid lines.  Secondary pathways of hydrocarbon and organic photolysis are shown as dashed lines.  Tertiary pathways of organic photolysis to form CO are shown as dotted lines.}
\label{fig:network}
\end{figure}

Based on this analysis, we conclude that the primary products of C$_2$H$_6$ + O are ethanol and acetaldehyde; of C$_2$H$_4$ + O are ethylene oxide and acetaldehyde; and of C$_2$H$_2$ + O is ketene.  Figure \ref{fig:network} summarizes this reaction network.  Reaction mechanisms will be discussed in detail in Section \ref{sec:disc_mech}.

\subsection{Product branching ratios}
\label{sec:br_ratio}
We derive branching ratios for the primary products formed from C$_2$H$_6$:$^{13}$CO$_2$ and C$_2$H$_4$:$^{13}$CO$_2$ experiments.  Since organic products are both formed and consumed over the course of the irradiation experiments, a branching ratio can be derived for either (i) the initial product distribution of a reaction, or (ii) for the steady-state product ratio.  The former is of direct use in astrochemical models, while the latter describes the end state of an experiment.  The steady-state branching ratio is found from the ratio of steady-state abundances $N_{ss}$ derived in growth curve fitting.  For the initial branching ratio, we use the growth curve fits to calculate the ratio of product yields at a very early fluence (0.1 $\times$ 10$^{17}$ ph cm$^{-2}$, or $\sim$ 3 minutes) using Equation \ref{eq:gc}.  This product yield at early times represents the formation-only regime, when repeated dissociations are not an issue (Section \ref{sec:model}).  In calculating the branching ratio, we include a band strength uncertainty of 20\% for ethanol and 50\% for ethylene oxide and acetaldehyde, the latter reflecting a higher uncertainty due to the use of gas-phase and theoretical values (Section \ref{sec:columns}).

\begin{deluxetable}{lcc} 
	\tabletypesize{\footnotesize}
	\tablecaption{Product branching ratios (\%) \label{tab:branch}}
	\tablecolumns{3} 
	\tablewidth{\textwidth} 
	\tablehead{
	       \colhead{} &
	       \colhead{C$_2$H$_6$} &
	       \colhead{C$_2$H$_4$} \\
		\colhead{}                 &
		\colhead{$\rm\frac{Eth}{Eth + Ac}$ }                  &
		\colhead{$\rm\frac{EtO}{EtO + Ac}$ }                  }
\startdata
Initial & 73.8$\pm$ 13.2 & 46.1 $\pm$ 12.1 \\
Steady-state & 36.9 $\pm$ 9.0 & 34.9 $\pm$ 7.8
\enddata
\tablecomments{Eth = ethanol, Ac = acetaldehyde, EtO = ethylene oxide.}
\end{deluxetable}

\begin{figure}
\centering
	\includegraphics[width=0.9\linewidth]{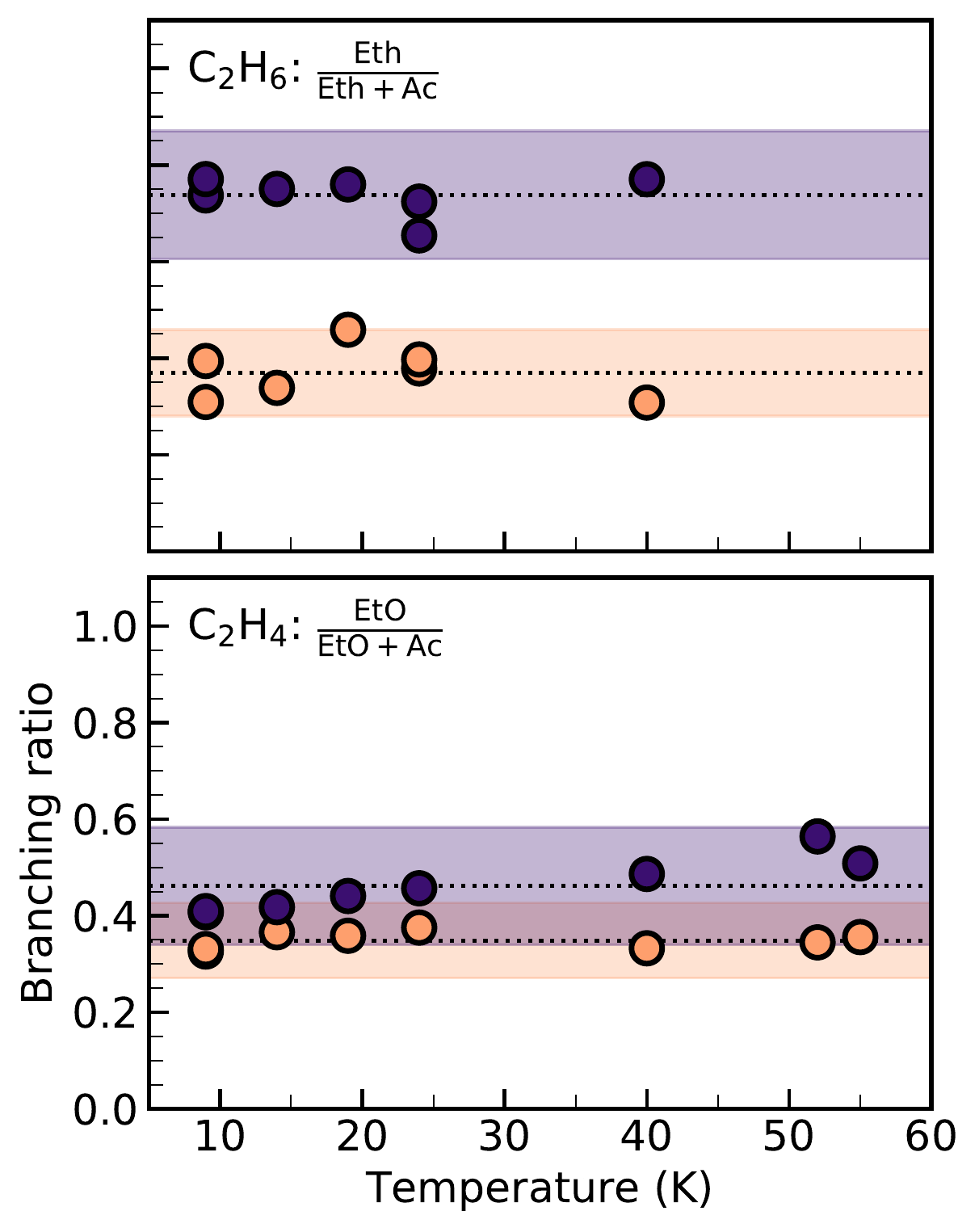}
	\caption{Product branching ratios for O atom reactions with C$_2$H$_6$ (top) and C$_2$H$_4$ (bottom) as a function of temperature.  Purple points show the initial branching ratio and orange points show the steady-state branching ratio.  Dotted lines and shaded regions show the average branching ratios across all temperatures and their uncertainties.  Eth = ethanol, Ac = acetaldehyde, EtO = ethylene oxide.}
\label{fig:ratios}
\end{figure}

Figure \ref{fig:ratios} shows the resulting branching ratios for all experiments as a function of temperature.  The branching ratio is represented in terms of the main oxygen atom addition product to each hydrocarbon.  We see no strong dependence of the branching ratio on temperature.  Table \ref{tab:branch} lists the average branching ratios across all temperatures.  We encourage astrochemical modelers to adopt the initial branching ratio, as this is representative of the statistical outcome of a single reaction.

\section{Discussion}
\label{sec:disc}

\subsection{Reaction Mechanisms} \label{sec:disc_mech}

\subsubsection{C$_2$H$_6$ + O} \label{sec:ethane_mech}

The primary products of oxygen atom reactions with ethane are ethanol and acetaldehyde.  Ethanol formation is readily explained by the insertion of an O($^1$D) atom into a C-H bond of ethane, as originally shown by \citet{DeMore1969} in liquid argon conditions.  This process is highly exothermic ($\sim$140 kcal), and in some cases the excited ethanol product may decompose prior to being collisionally stabilized in the ice matrix.  CH$_3$CHO + H$_2$ is the most stable product of ethanol fragmentation \citep[][and references therein]{Shu2001}, and CH$_3$CHO observed at early times in our experiments is likely formed as a direct product of oxygen insertion into ethane in this way.  This is analogous to the formation of methanol and formaldehyde as the primary products of O($^1$D) reactions with methane \citep{Bergner2017}.

In addition to this first-generation pathway to acetaldehyde, there is likely a second-generation pathway of e.g. ethanol photolysis.  As seen in Figure \ref{fig:vsgc}, ethanol production plateaus towards the end of an experiment, while acetaldehyde does not appear to have reached a steady state.  The acetaldehyde growth curve therefore consists of first-generation formation at early times, and an increasing contribution of second-generation chemistry at later times. 

We observe no temperature dependence to the ethanol or acetaldehyde formation rate constants from 9 -- 40 K (Figure \ref{fig:arrh}), supporting that O($^1$D) atom insertions into a C-H bond of ethane proceed with a negligible energy barrier.  We note that oxygen atoms generated by CO$_2$ photolysis may be born with excess energy, and so we cannot rule out that a hot atom mechanism is masking a barrier.  However, taken together with gas-phase studies showing no temperature dependence to O($^1$D) insertions into C$_2$H$_6$ \citep{Nunez2018}, a barrierless process is a likely explanation for our observed results.

\subsubsection{C$_2$H$_4$ + O} \label{sec:ethylene_mech}

In our experiments, the primary products of C$_2$H$_4$ + O are ethylene oxide and acetaldehyde.  \citet{DeMore1969} has demonstrated that both species can be formed from either O($^1$D) or O($^3$P) reactions with C$_2$H$_4$ in liquid argon, and it is therefore possible that both electronic states contribute to the chemistry that we observe. Indeed, the temperature dependence shown in Figure \ref{fig:arrh} suggests that two different processes are at play: a slower but barrierless channel accessible at low temperatures, and a barriered channel accessible above $\sim$20 K.  A likely explanation for this behavior is if O($^1$D) and O($^3$P) are responsible for the two pathways.  \citet{Ward2011} performed experiments with thermal O($^3$P) atoms + C$_2$H$_4$ at temperatures from 12--90 K, and found that ethylene oxide and acetaldehyde formation could be modeled using a Langmuir-Hinshelwood channel that becomes active around 20--30 K, perfectly consistent with the temperature dependence that we observe.  The 190 $\pm$ 45 K barrier determined for the LH mechanism in \citet{Ward2011} is somewhat higher than the $\sim$60--80 K barriers derived from our experiments (Section \ref{sec:arrh}); this may be related to the very different methods for measuring energy barriers used in both studies.  We note that a hot atom effect is not likely to be responsible for this difference, since O($^3$P) is not a direct photolysis product, and O($^1$D) should thermalize many orders of magnitude faster than it relaxes to O($^3$P) (see Section \ref{sec:acetylene_mech})

We note that additional processes besides an O($^3$P) contribution could also explain the temperature trend we observe.  A diffusion-limited pathway involving O($^1$D) could become important at warmer temperatures, though we would then expect to see a similar trend for the other hydrocarbons.  Alternatively, a tunneling contribution could produce a flattening of the reaction rate at low temperatures.  While it has been shown that tunneling may be important to O atom diffusion on ice surfaces \citep{Minissale2013}, to date there is no example of O atom tunneling playing an important role in reaction rates.  In the one case where it has been theoretically tested, the CO + O $\rightarrow$ CO$_2$ reaction, tunneling was found to contribute to the reaction rate at temperatures below $\sim$80 K, however the absolute rate is so slow that it is unlikely to be an important pathway \citep{Goumans2010}.

Most reactions between O + C$_2$H$_4$ likely proceed via an oxygen addition mechanism.  O($^3$P) and O($^1$D) can both add to the ethylene double bond to form the $\mathrm{\dot{C}H_2CH_2\dot{O}}$ biradical, followed by either ring closure to form ethylene oxide, or rearrangement to form acetaldehyde \citep{Hawkins1983}.  Initially formed ethylene oxide may also undergo isomerization to acetaldehyde if it is produced with sufficient energy \citep{DeMore1969}.

Besides an O addition pathway, O($^1$D) atoms could insert into C-H bonds of ethylene, though this is likely a minor channel:  \citet{DeMore1969} has shown that O($^1$D) atoms attack an ethylene C=C bond five times faster than inserting into a C-H bond.  Though the insertion product vinyl alcohol has been previously demonstrated to form from C$_2$H$_4$ + O($^1$D) by \citet{Hawkins1983}, we do not detect it in our experiments.  If vinyl alcohol is forming, it may be a transient product: vinyl alcohol can undergo tautomerization to form acetaldehyde with a moderate energy barrier \citep[85 kcal/mol;][]{Bouma1977}, which may be surmounted by the exothermicity of its formation.  The thin ices used in our experiments also limit our ability to detect small yields of vinyl alcohol.  Thus, our non-detection of vinyl alcohol may be explained by a tendency to convert to acetaldehyde or other products, combined with sensitivity limitations.

\subsubsection{C$_2$H$_2$ + O} \label{sec:acetylene_mech}

Ketene is the main organic product resulting from O + C$_2$H$_2$ in our experiments.  Theory predicts that both O($^3$P) and O($^1$D) can add to the C$\equiv$C bond, followed by a rearrangement to form ketene \citep{Rajak2014}.  O($^3$P) addition to C$_2$H$_2$ proceeds with a theoretical barrier of at least 1760 K, whereas O($^1$D) addition is predicted to be barrierless \citep{Nguyen2005, Rajak2014}.  Gas phase O($^1$D) + C$_2$H$_2$ reaction rates show no temperature dependence \citep{Nunez2018}, confirming the theoretical prediction of a barrierless chemistry.  In our experiments, ketene formation rate constants show no temperature dependence from 9--52 K, consistent with a barrierless O($^1$D) mechanism.

O($^3$P) likely does not play an important role in ketene formation under our experimental conditions.  \citet{Haller1962} demonstrated ketene formation by O($^3$P) + C$_2$H$_2$ in argon matrices at 20 K; however, in their experiments O($^3$P) is formed directly from N$_2$O photolysis, and a hot atom mechanism is active.  In our experiments, O($^3$P) is not a direct photolysis product, but rather forms only upon O($^1$D) relaxation.  A hot O($^3$P) mechanism would require the O($^1$D) thermalization timescale to be longer than its lifetime.  While the thermalization of hot oxygen atoms is not well characterized, studies of photo-produced hot OH radicals indicate that the thermalization timescale is very fast, on the order of picoseconds \citep{Andersson2006, Andersson2008}.  Given that this is many orders of magnitude faster than the O($^1$D) lifetime of seconds, we do not expect hot O($^3$P) atoms to be present in our experiments.  Since we do not observe a temperature dependence in our Arrhenius plots, and we have ruled out a hot atom contribution, we conclude that the barriered O($^3$P) + C$_2$H$_2$ reaction is not an important contributor to the observed chemistry, and that most of the ketene formation observed in our experiments is due to O($^1$D) chemistry.

Ethynol, the direct insertion product of O($^1$D) + acetylene, is not observed in the IR spectrum in our experiments \citep{Hochstrasser1989}.  Theoretical studies show that ethynol could convert to formylcarbene and then to ketene \citep{Girard2003}, and so if an insertion chemistry is active it could be contributing to the observed ketene formation.  However, we expect that the overall contribution will be minor since, as with ethylene, addition to the multiple bond should occur faster than insertion.

\vspace{0.1in}
\subsubsection{Radical recombination vs. direct reaction}
All of the products that are observed in these experiments can be readily explained by a reaction scheme relying on direct addition of oxygen atoms to hydrocarbons.  We rule out a radical recombination mechanism for the main O-containing organics based on the following lines of reasoning.  First, oxygen-bearing organic products are not observed in the control experiments of CO:hydrocarbon mixtures, indicating that reactions of CO with hydrocarbon photodissociation products do not contribute to organic product formation.  Moreover, we do not observe $^{13}$C-labeled O-containing organic products, indicating that the $^{12}$C hydrocarbon backbone is preserved throughout this chemistry.  We also do not detect the radical species CH$_3$, H$^{13}$CO, or HCO in the infrared, finding upper limits of $\sim$0.2ML CH$_3$ and $\sim$0.1ML H$^{13}$CO and HCO.  Other products signifying an active radical chemistry are not detected, including \textit{cis-}HO$^{13}$CO and \textit{trans-}HO$^{13}$CO at 1756 and 1792 cm$^{-1}$, respectively \citep{Milligan1971}; polyethylene at 2925 cm$^{-1}$ \citep{Krimm1956}; diacetylene at 3280 cm$^{-1}$ \citep{Freund1965}; and propanoic acid at 1776 cm$^{-1}$ \citep{Macoas2005}.  We thus conclude that radical recombinations do not play an important role in the observed organic product formation, and direct reaction of oxygen atoms with the hydrocarbons is responsible.  Again, Figure \ref{fig:network} summarizes how the reactants and products are connected by oxygen atom driven chemistry, consistent with the mechanisms for organic product formation described in Sections \ref{sec:ethane_mech} -- \ref{sec:acetylene_mech}.

\subsection{Branching ratios comparison}
\label{sec:br_disc}

We now discuss the branching ratios derived in Section \ref{sec:br_ratio} in light of the reaction mechanisms presented in Section \ref{sec:disc_mech}.  For O + C$_2$H$_6$, the initial branching ratio is 74\% ethanol and 26\% acetaldehyde.  At early times both products are expected to result directly from O($^1$D) insertions, with acetaldehyde forming when the excited ethanol product is not stabilized (Section \ref{sec:ethane_mech}). Based on the derived branching ratio, in most cases the excited ethanol product survives intact.  The branching ratio to ethanol from ethane is somewhat higher than the branching ratio to methanol from methane \citep[$\sim$65\%; ][]{Bergner2017}.  If confirmed to be significant, this difference suggests that stabilization of the excited insertion product is more efficient for larger molecules.

The initial branching ratio for O + C$_2$H$_4$ is 47\% ethylene oxide and 53\% acetaldehyde.  Both products are expected to form from the $\mathrm{\dot{C}H_2CH_2\dot{O}}$ biradical (Section \ref{sec:ethylene_mech}).  The parity in the branching ratio suggests that there is no strong preference for ring closure verses rearrangement from this shared intermediate.  We note also that there is no strong temperature dependence to the branching ratio (Figure \ref{fig:ratios}) despite evidence for an increasing contribution of O($^3$P) at higher temperatures (Figure \ref{fig:arrh}).  This indicates that the distribution of products from this chemistry is insensitive to the electronic state of the oxygen atom.

The steady-state branching ratios for ethanol from ethane and ethylene oxide from ethylene are both $\sim$35\%.  In both cases, second-generation chemistry increases the relative yield of acetaldehyde, though this is much more important in the case of O + C$_2$H$_6$ chemistry compared to O + C$_2$H$_4$ chemistry.

The branching ratios that we derive are comparable to those found in \citet{DeMore1969}, whose reactions took place at 87 K and diluted in liquid argon.  For C$_2$H$_6$ + O($^1$D), the authors find that $\sim$75\% of the decomposed O$_3$ goes to ethanol and 10\% to acetaldehyde, a relative branching ratio of 88\% ethanol.  For C$_2$H$_4$ + O($^1$D), they find that 10\% of the decomposed O$_3$ goes to ethylene oxide and 22\% to acetaldehyde, or a relative branching of 31\% ethylene oxide.  While we find a slightly lower ethanol yield (74\%) and higher ethylene oxide yield (47\%), it is not clear if these differences are significant given the uncertainties in our branching ratios.  Better constraints on the product band strengths are required to resolve whether there is a real difference in the branching ratios under our experimental conditions.

\subsection{Astrophysical implications}
We have demonstrated a low-temperature formation pathway to the organic molecules acetaldehyde, ethanol, ethylene oxide, and ketene via oxygen atom reactions with the hydrocarbons ethane, ethylene, and acetylene.  As mentioned previously, excited O($^1$D) atoms should be readily generated in ISM ices by the photolysis or radiolysis of oxygen-bearing molecules like H$_2$O and CO$_2$.  Because thermal diffusion is accessible to atoms at lower temperatures than for larger radical fragments, oxygen atoms should still be able to scan the dust grain and encounter co-reactants even in very cold ISM regions.  

We observe a temperature-independent component to the reaction rates of oxygen with all three hydrocarbons.  While we cannot rule out the presence of a small energy barrier masked by a hot atom mechanism, gas-phase and theoretical studies point to a barrierless mechanism involving the O($^1$D) atom.  Excited oxygen atoms must be generated in situ in ISM ices by photolysis or radiolysis of oxygen-bearing molecules, and therefore we expect the overall mechanism in the ISM to be analogous to our experimental scheme.  All this points to an energetically feasible chemistry even at very low temperatures, as long as the ice is exposed to some dissociative radiation.  Additional chemistry involving the O($^3$P) atom may be also become energetically accessible at warmer temperatures.

Acetaldehyde and ketene have been commonly detected towards very cold ISM regions like pre-stellar cores and embedded protostars, with rotational temperatures of 5--10 K \citep{Oberg2010, Bacmann2012, Bergner2017b}.  Oxygen atom chemistry could certainly contribute to acetaldehyde and ketene formation under such conditions, where diffusion of larger radical species is not expected to be efficient.  Because acetaldehyde is the main steady-state product of oxygen atom reactions with both ethane and ethylene, we expect its formation to be particularly efficient.  As with any products of low-temperature grain-surface chemistry, the observed gas-phase abundances may be explained by non-thermal desorption mechanisms such as chemical desorption or photodesorption.

Ethylene oxide has been detected towards several massive star-forming regions with moderate rotational temperatures of $\sim$10--30 K and a range of column density ratios respect to acetaldehyde of 0.07--0.8 \citep{Ikeda2001}.  It was recently detected for the first time towards a low-mass protostar, with a rotational temperature of $\sim$125 K and a column density ratio of 0.09 with respect to acetaldehyde \citep{Lykke2017}.  The branching ratio we derive corresponds to a c-C$_2$H$_4$O/CH$_3$CHO ratio of about unity.  To our knowledge, oxygen atom reaction with C$_2$H$_4$ is the only mechanism to ethylene oxide formation available in the literature, whereas acetaldehyde formation can also be explained by radical recombination or hydrogenation chemistry, or indeed oxygen atom reactions with ethane.  Thus, it is reasonable to expect that that the ethylene oxide to acetaldehyde ratio is diluted as other chemical pathways to acetaldehyde become active.

Ethanol has been detected mainly in warm interstellar environments, with rotational temperatures typically around or above 100 K \citep{Ikeda2001, Bisschop2007,Fuente2014}.  It is unlikely that oxygen atom reactions play an important role under these conditions, and indeed the observed ratios with respect to acetaldehyde range from a factor of 4 to several hundred, clearly at odds with our derived ratio of order unity.  While it may be difficult to detect ethanol in colder environments given its high sublimation temperature, constraining the ratio in cold environments would be a useful test of the O atom chemistry presented here.

\section{Conclusions}
Based on our experiments of oxygen atom reactions with the C$_2$ hydrocarbons ethane, ethylene, and acetylene under ISM-like conditions, we conclude the following:

\begin{enumerate}
\item Oxygen atoms react with ethane to form ethanol and acetaldehyde, with a branching ratio to ethanol of 0.74; with ethylene to form ethylene oxide and acetaldehyde, with a branching ratio to ethylene oxide of 0.47; and with acetylene to form ketene.
\item For ethane and acetylene, product formation rate constants are temperature independent for the entire range of temperatures studied.  For ethylene, product formation rate constants show no temperature dependence below 25 K, but an increase with temperature above this.  Taken together with literature studies, this suggests that all three hydrocarbons react with O($^1$D) with no effective energy barrier, while a barriered reaction of ethylene with O($^3$P) becomes important at warmer temperatures.
\item Unlike in the gas phase, the solid-state organic products of oxygen atom reactions with hydrocarbons are stabilized rather than undergoing unimolecular dissociation to form radicals.
\item Oxygen atom chemistry is energetically feasible under ISM-like conditions and leads to the formation of COMs from hydrocarbons.  In particular, this chemistry could help to explain detections of acetaldehyde and ketene at $<$10 K during the earliest stages of star formation.
\end{enumerate}

\noindent The authors thank I. Cooke, R. Martin-Domenech, P. Maksiutenko, and L. Kelley for valuable feedback.  J.B.B acknowledges funding from the National Science Foundation Graduate Research Fellowship under grant DGE1144152.  This work was supported by an award from the Simons Foundation (SCOL \# 321183, KO).

\software{
{\fontfamily{qcr}\selectfont NumPy} \citep{VanDerWalt2011},
{\fontfamily{qcr}\selectfont Matplotlib} \citep{Hunter2007},
{\fontfamily{qcr}\selectfont emcee} \citep{Foreman-Mackey2013}
}

\clearpage
\appendix
\FloatBarrier

\begin{figure*}
	\includegraphics[width=\linewidth]{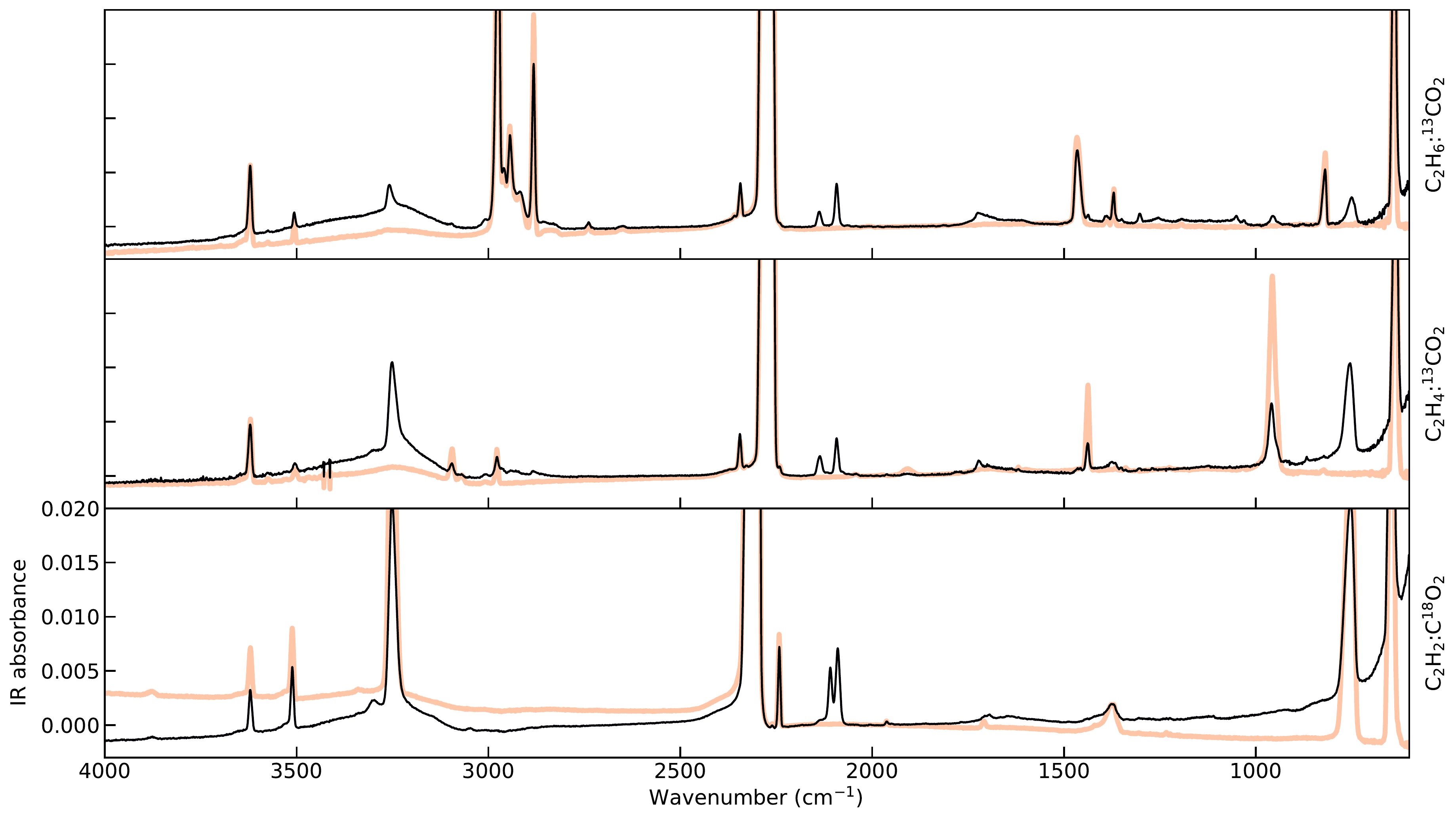}
	\caption{Full IR spectra for 9 K irradiation experiments of a C$_2$H$_6$:$^{13}$CO$_2$ mixture (top), C$_2$H$_4$:$^{13}$CO$_2$ mixture (middle), and C$_2$H$_2$:C$^{18}$O$_2$ mixture (bottom).  Orange lines show the pre-irradiation spectra, and solid black lines show the spectra following $\sim$2h irradiation.}
\label{fig:ir_full}
\end{figure*}

\begin{figure}
\plottwo{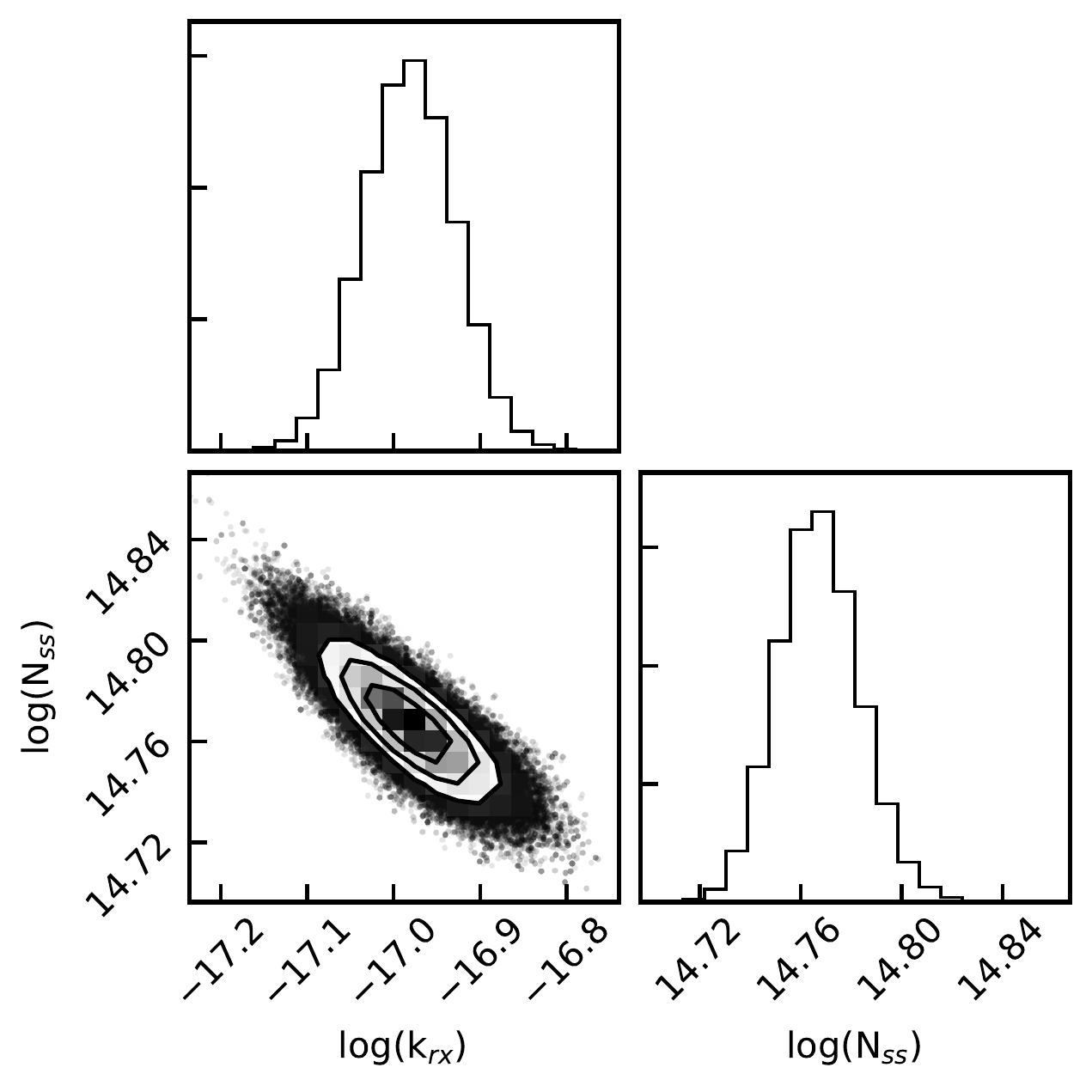}{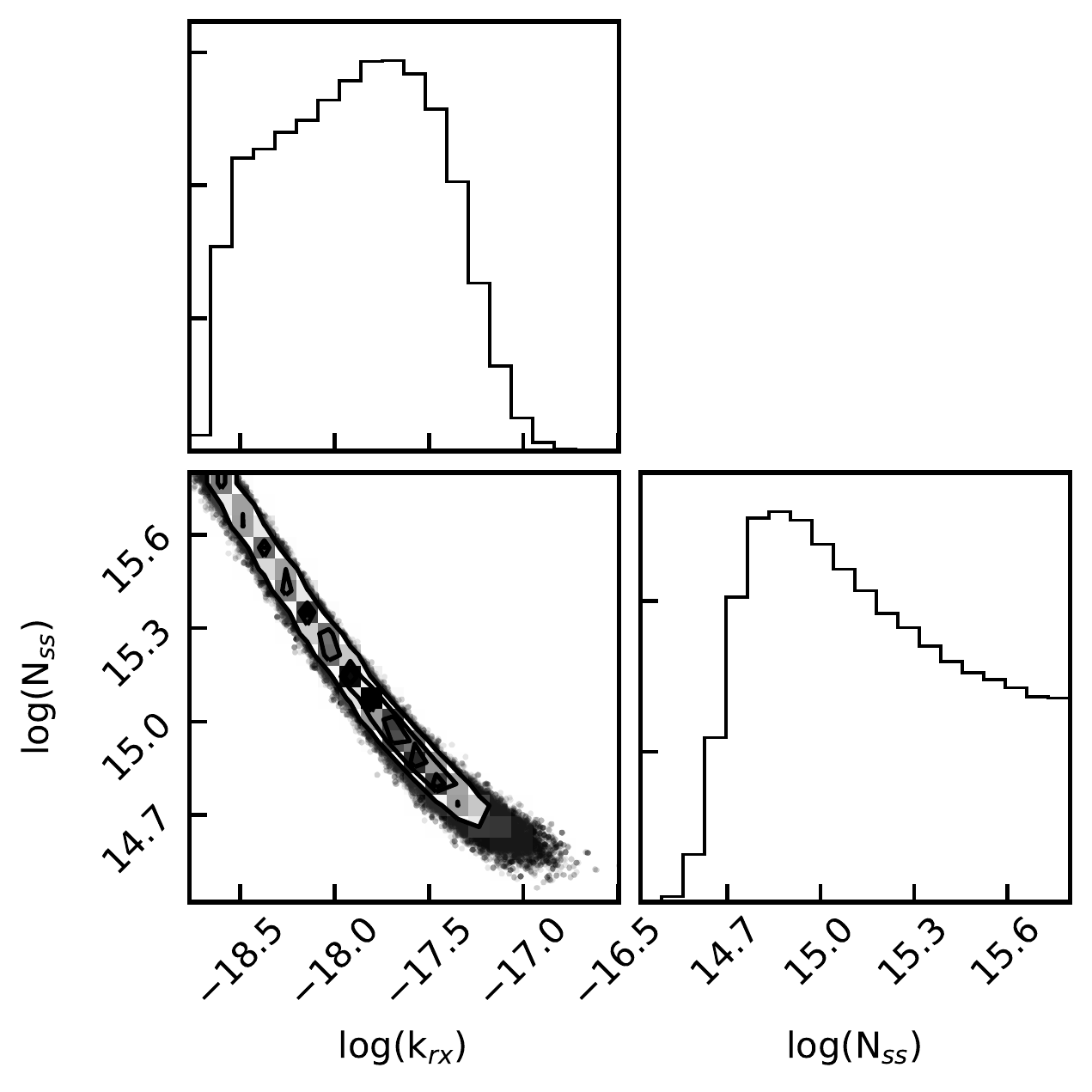}
\caption{Example corner plots of growth curve fitting for the C$_2$H$_6$:$^{13}$CO$_2$ irradiation products ethanol (left) and acetaldehyde (right).}
\label{fig:corner_c2h6}
\end{figure}

\begin{figure}
\plottwo{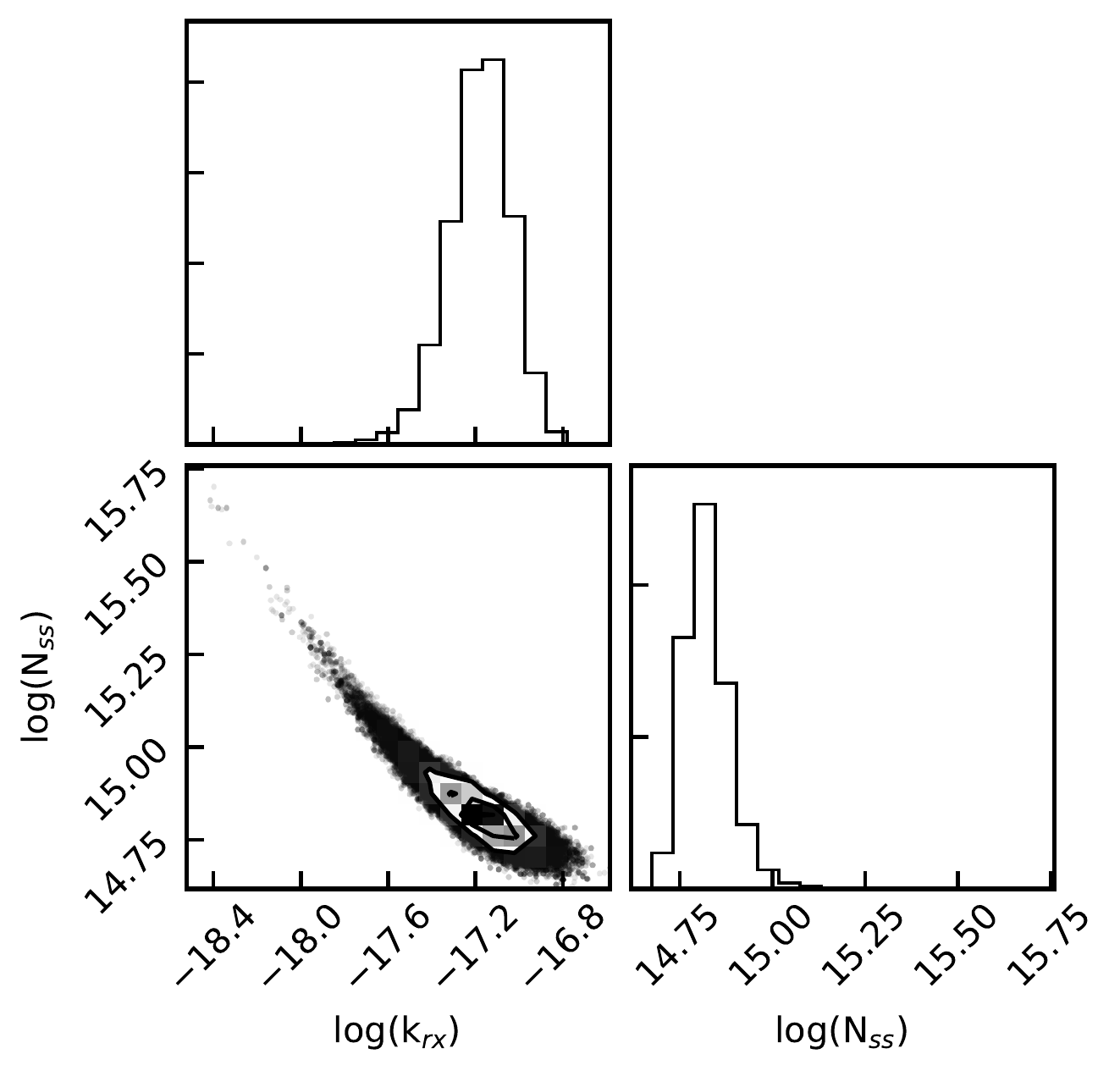}{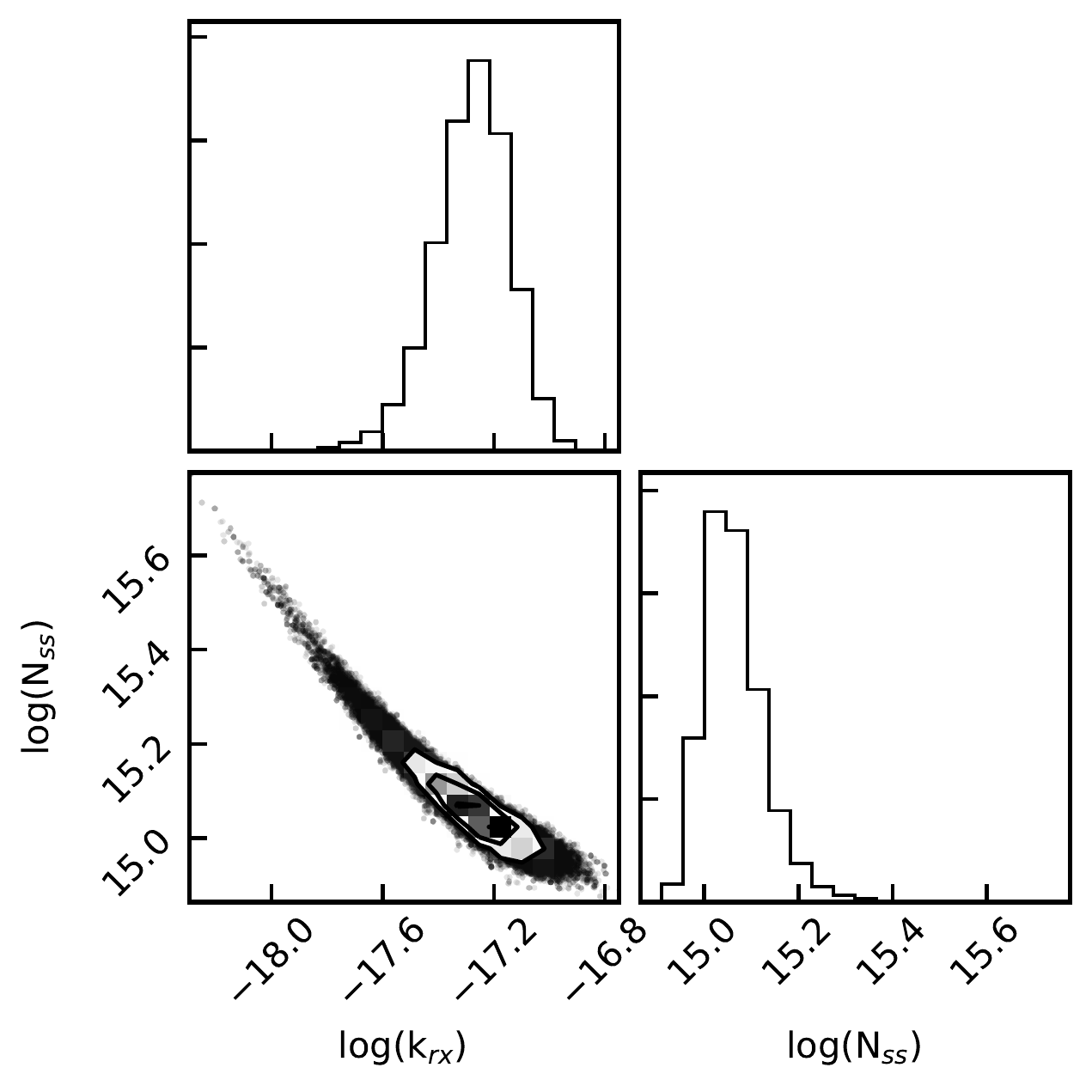}
\caption{Example corner plots of growth curve fitting for the C$_2$H$_4$:$^{13}$CO$_2$ irradiation products ethylene oxide (left) and acetaldehyde (right).}
\label{fig:corner_c2h4}
\end{figure}

\begin{figure}
\includegraphics[width=0.45\linewidth]{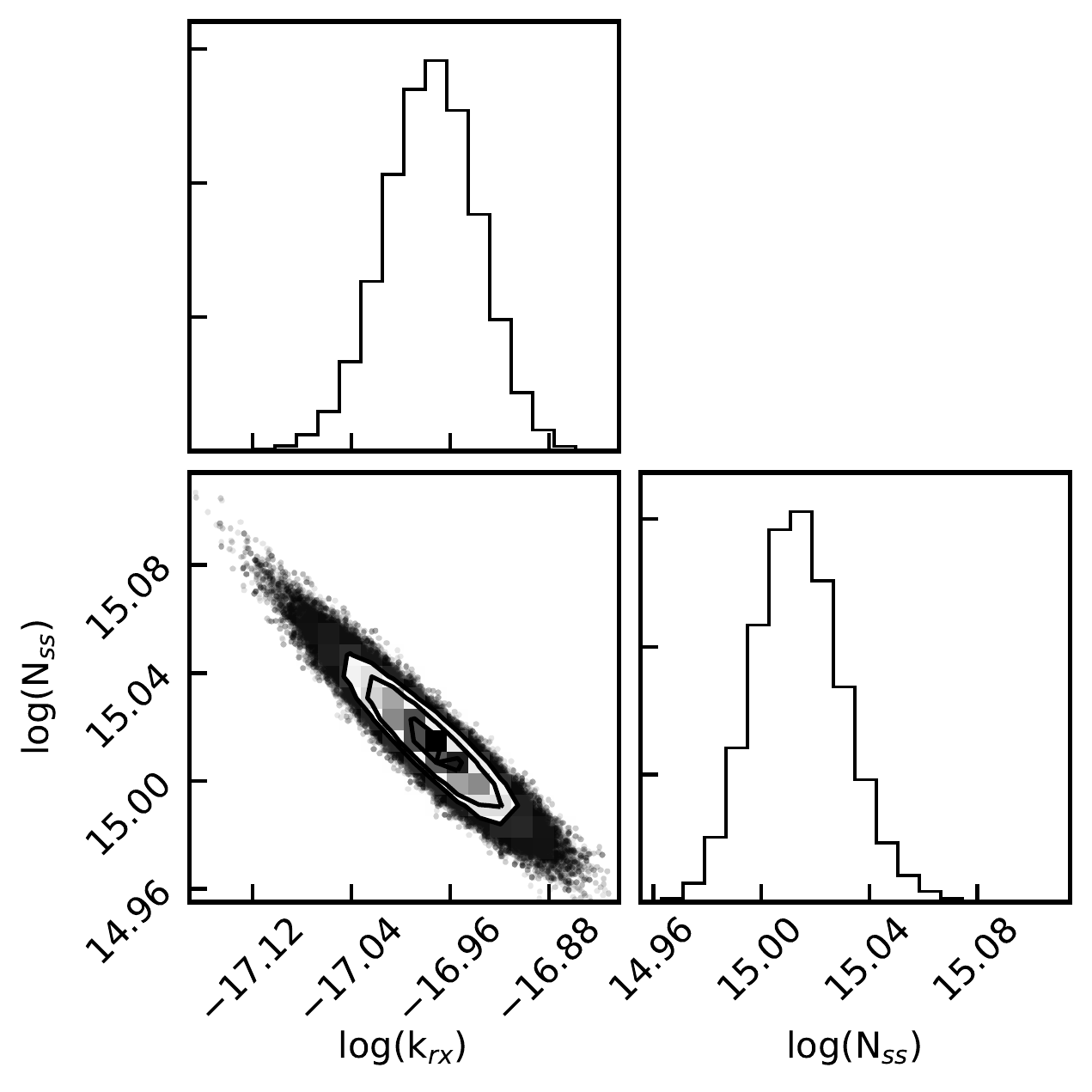}
\caption{Example corner plots of growth curve fitting for the C$_2$H$_2$:C$^{18}$O$_2$ irradiation product CH$_2$C$^{18}$O.}
\label{fig:corner_c2h2}
\end{figure}

\begin{figure}
	\includegraphics[width=0.9\linewidth]{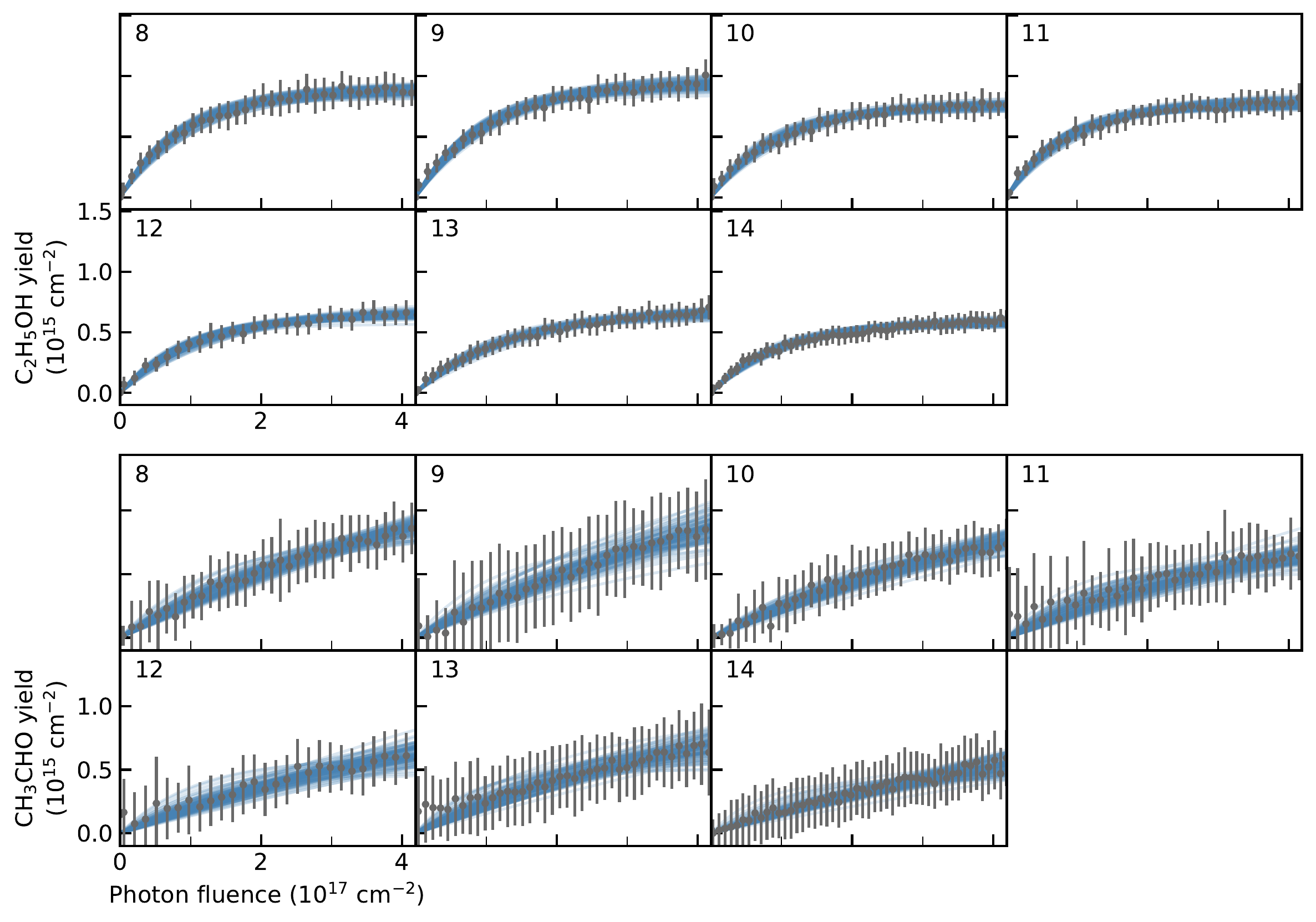}
	\caption{Growth curves with kinetic model fits for the main organic products of C$_2$H$_6$:$^{13}$CO$_2$ irradiations.  Experimental data is shown as grey points, and blue lines show fits drawn from the posterior probability distribution.  Experiment numbers are listed in the upper left corner of each panel.}
\label{fig:gc_c2h6}
\end{figure}

\begin{figure}
	\includegraphics[width=0.9\linewidth]{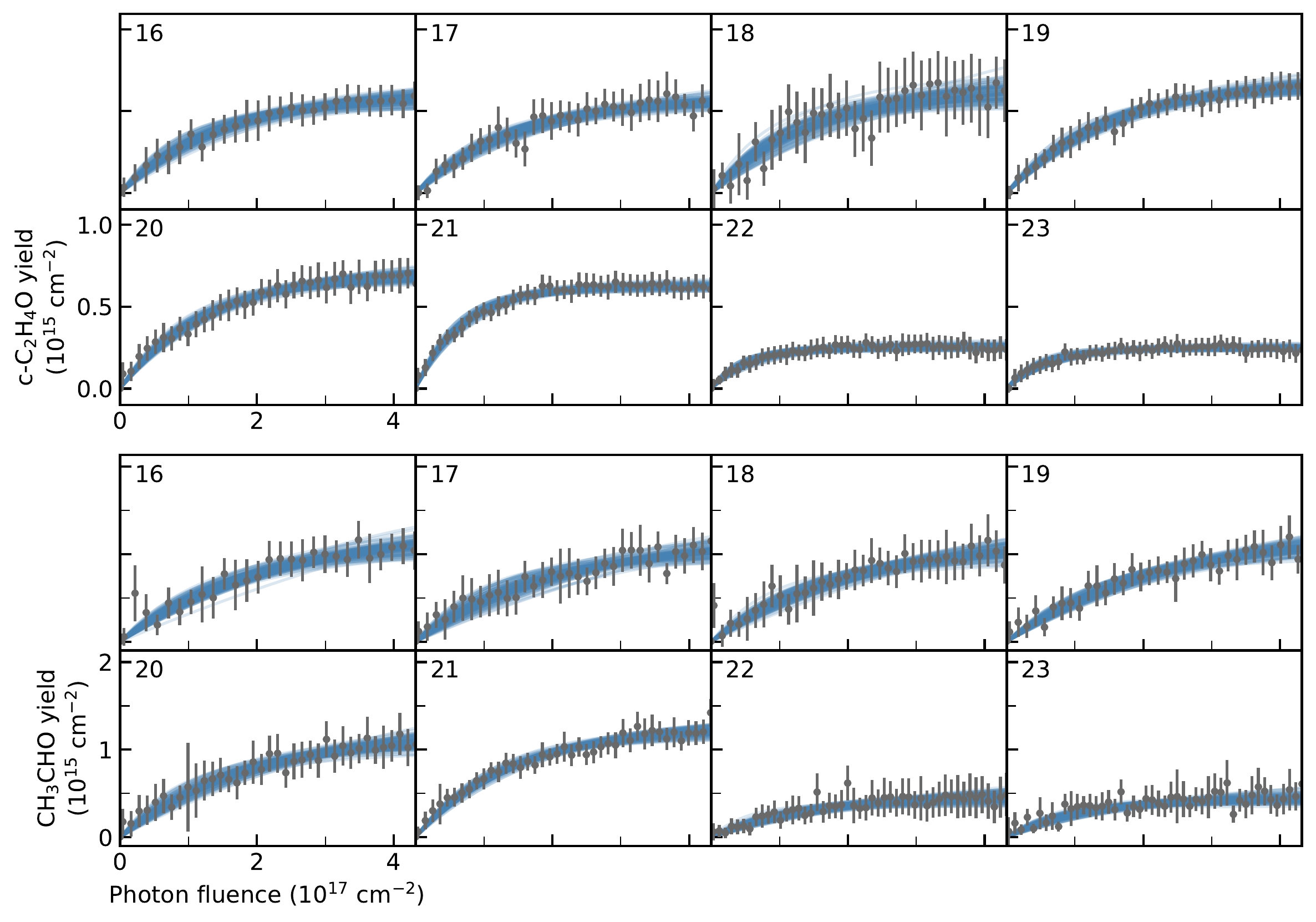}
	\caption{Growth curves with kinetic model fits for the main organic products of C$_2$H$_4$:$^{13}$CO$_2$ irradiations.  Experimental data is shown as grey points, and blue lines show fits drawn from the posterior probability distribution.  Experiment numbers are listed in the upper left corner of each panel.}
\label{fig:gc_c2h4}
\end{figure}

\begin{figure}
	\includegraphics[width=0.9\linewidth]{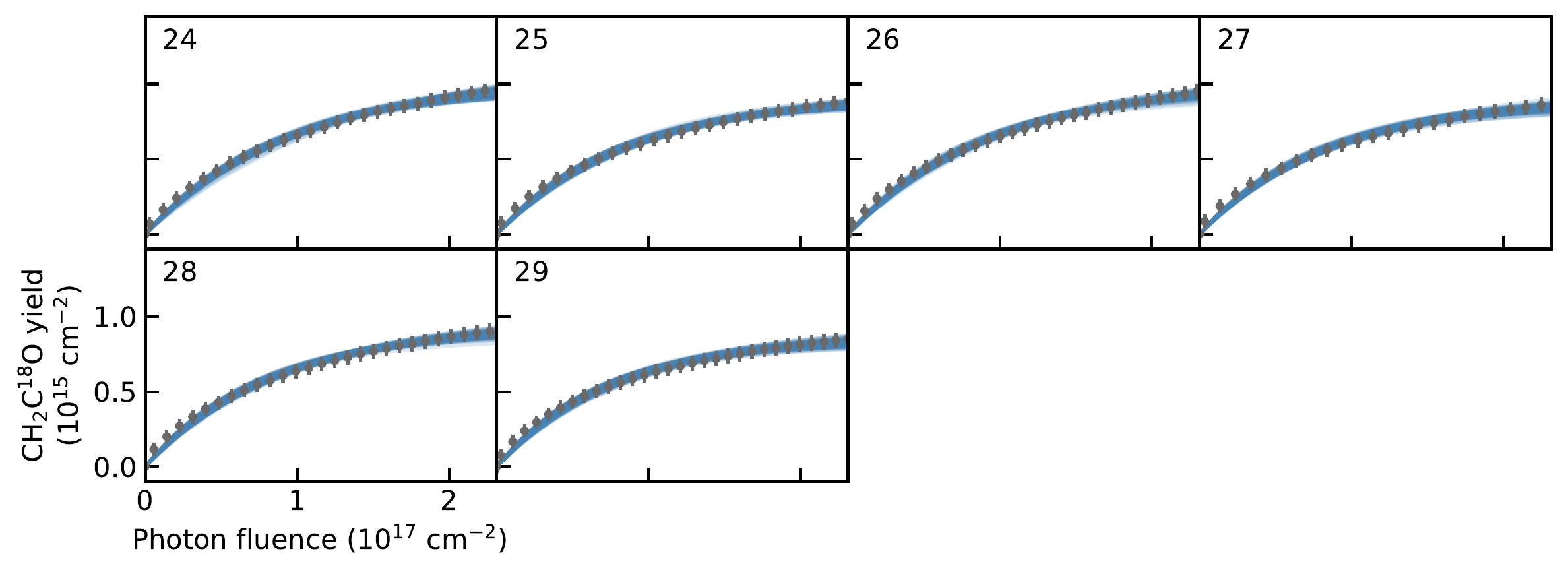}
	\caption{Growth curves with kinetic model fits for the main organic product of C$_2$H$_2$:C$^{18}$O$_2$ irradiations.  Experimental data is shown as grey points, and blue lines show fits drawn from the posterior probability distribution.  Experiment numbers are listed in the upper left corner of each panel.}
\label{fig:gc_c2h2}
\end{figure}

Figure \ref{fig:ir_full} shows the pre- and post-irradiation spectra for each hydrocarbon:CO$_2$ ice mixture.  Table \ref{tab:gcfits} lists the best-fit parameters and uncertainties resulting from growth curve fitting of the main organic products of C$_2$H$_6$:$^{13}$CO$_2$, C$_2$H$_4$:$^{13}$CO$_2$, and C$_2$H$_2$:C$^{18}$O$_2$ irradiations.  
Figures \ref{fig:corner_c2h6} -- \ref{fig:corner_c2h2} are example corner plots showing parameter degeneracies for the kinetic fits.
Figures \ref{fig:gc_c2h6} -- \ref{fig:gc_c2h2} show all experimental growth curves as well as draws from the posterior probability distribution of the mcmc fitting.  

\begin{deluxetable*}{lllllcll} 
	\tabletypesize{\footnotesize}
	\tablecaption{Growth curve best-fit parameters \label{tab:gcfits}}
	\tablecolumns{6} 
	\tablewidth{\textwidth} 
	\tablehead{
		\colhead{Exp.}                &
		\colhead{$k_{rxn}$}       &
		\colhead{N$_{ss}$}        &
		\colhead{|}                     &
		\colhead{$k_{rxn}$}        &
		\colhead{N$_{ss}$}         \\
		\colhead{}                                       &
		\colhead{(ph$^{-1}$)} &
		\colhead{(10$^{15}$ cm$^{-2}$)}     &
		\colhead{ }                                       &
		\colhead{(ph$^{-1}$)} &
		\colhead{(10$^{15}$ cm$^{-2}$)}      }
\startdata
& \multicolumn{2}{c}{C$_2$H$_5$OH} &&  \multicolumn{2}{c}{CH$_3$CHO} \\
\hline
8 & 1.0 $^{+ 0.1}_{- 0.1}$ x 10$^{-17}$ & 0.90 $^{+ 0.04}_{- 0.03}$ & & 2.3 $^{+ 1.3}_{- 1.7}$ x 10$^{-18}$ & 1.38 $^{+ 1.21}_{- 0.38}$\\
9 & 9.9 $^{+ 1.1}_{- 1.2}$ x 10$^{-18}$ & 0.94 $^{+ 0.04}_{- 0.03}$ & & 1.3 $^{+ 0.7}_{- 1.8}$ x 10$^{-18}$ & 2.10 $^{+ 2.19}_{- 0.97}$\\
10 & 1.1 $^{+ 0.1}_{- 0.2}$ x 10$^{-17}$ & 0.77 $^{+ 0.03}_{- 0.03}$ & & 1.7 $^{+ 1.0}_{- 1.5}$ x 10$^{-18}$ & 1.50 $^{+ 1.56}_{- 0.50}$\\
11 & 1.2 $^{+ 0.1}_{- 0.2}$ x 10$^{-17}$ & 0.78 $^{+ 0.03}_{- 0.03}$ & & 3.0 $^{+ 2.0}_{- 2.9}$ x 10$^{-18}$ & 0.92 $^{+ 1.11}_{- 0.25}$\\
12 & 9.1 $^{+ 1.3}_{- 1.6}$ x 10$^{-18}$ & 0.66 $^{+ 0.04}_{- 0.03}$ & & 2.0 $^{+ 1.4}_{- 2.7}$ x 10$^{-18}$ & 1.09 $^{+ 1.77}_{- 0.42}$\\
13 & 7.6 $^{+ 1.0}_{- 1.1}$ x 10$^{-18}$ & 0.69 $^{+ 0.04}_{- 0.03}$ & & 2.6 $^{+ 1.7}_{- 2.8}$ x 10$^{-18}$ & 1.04 $^{+ 1.41}_{- 0.34}$\\
14 & 1.0 $^{+ 0.1}_{- 0.1}$ x 10$^{-17}$ & 0.58 $^{+ 0.02}_{- 0.02}$ & & 1.3 $^{+ 0.9}_{- 2.3}$ x 10$^{-18}$ & 1.31 $^{+ 2.21}_{- 0.67}$\\
\hline
  & \multicolumn{2}{c}{c-C$_2$H$_4$O} & & \multicolumn{2}{c}{CH$_3$CHO} \\
\hline
16 & 6.8 $^{+ 1.5}_{- 1.8}$ x 10$^{-18}$ & 0.61 $^{+ 0.06}_{- 0.04}$ & & 4.7 $^{+ 1.4}_{- 1.5}$ x 10$^{-18}$ & 1.27 $^{+ 0.24}_{- 0.15}$\\
17 & 7.0 $^{+ 1.3}_{- 1.4}$ x 10$^{-18}$ & 0.58 $^{+ 0.05}_{- 0.04}$ & & 4.9 $^{+ 1.5}_{- 1.8}$ x 10$^{-18}$ & 1.18 $^{+ 0.24}_{- 0.14}$\\
18 & 6.7 $^{+ 1.9}_{- 2.3}$ x 10$^{-18}$ & 0.66 $^{+ 0.10}_{- 0.07}$ & & 5.4 $^{+ 1.5}_{- 1.7}$ x 10$^{-18}$ & 1.14 $^{+ 0.19}_{- 0.13}$\\
19 & 6.8 $^{+ 1.0}_{- 1.2}$ x 10$^{-18}$ & 0.68 $^{+ 0.04}_{- 0.04}$ & & 4.7 $^{+ 1.1}_{- 1.2}$ x 10$^{-18}$ & 1.22 $^{+ 0.18}_{- 0.12}$\\
20 & 8.1 $^{+ 1.1}_{- 1.3}$ x 10$^{-18}$ & 0.71 $^{+ 0.04}_{- 0.03}$ & & 5.7 $^{+ 1.3}_{- 1.5}$ x 10$^{-18}$ & 1.18 $^{+ 0.15}_{- 0.10}$\\
21 & 1.4 $^{+ 0.1}_{- 0.1}$ x 10$^{-17}$ & 0.63 $^{+ 0.01}_{- 0.01}$ & & 7.3 $^{+ 0.8}_{- 0.9}$ x 10$^{-18}$ & 1.26 $^{+ 0.06}_{- 0.05}$\\
22 & 1.8 $^{+ 0.3}_{- 0.4}$ x 10$^{-17}$ & 0.25 $^{+ 0.01}_{- 0.01}$ & & 7.1 $^{+ 2.1}_{- 2.6}$ x 10$^{-18}$ & 0.48 $^{+ 0.09}_{- 0.06}$\\
23 & 1.8 $^{+ 0.3}_{- 0.5}$ x 10$^{-17}$ & 0.25 $^{+ 0.01}_{- 0.01}$ & & 9.0 $^{+ 2.3}_{- 2.9}$ x 10$^{-18}$ & 0.45 $^{+ 0.06}_{- 0.04}$\\
\hline
  & \multicolumn{2}{c}{CH$_2$C$^{18}$O} & & \multicolumn{2}{c}{ } \\
\hline
24 & 1.1 $^{+ 0.1}_{- 0.1}$ x 10$^{-17}$ & 1.03 $^{+ 0.04}_{- 0.04}$ & \\
25 & 1.2 $^{+ 0.1}_{- 0.1}$ x 10$^{-17}$ & 0.92 $^{+ 0.04}_{- 0.03}$ & \\
26 & 1.1 $^{+ 0.1}_{- 0.1}$ x 10$^{-17}$ & 1.00 $^{+ 0.04}_{- 0.03}$ & \\
27 & 1.2 $^{+ 0.1}_{- 0.1}$ x 10$^{-17}$ & 0.89 $^{+ 0.04}_{- 0.03}$ & \\
28 & 1.2 $^{+ 0.1}_{- 0.1}$ x 10$^{-17}$ & 0.94 $^{+ 0.04}_{- 0.03}$ & \\
29 & 1.3 $^{+ 0.1}_{- 0.1}$ x 10$^{-17}$ & 0.86 $^{+ 0.03}_{- 0.03}$ & \\
\enddata
\end{deluxetable*}

\end{document}